\documentclass[10pt,twocolumn,twoside]{IEEEtran}
\pdfoutput=1
\usepackage{etex}

\usepackage[utf8]{inputenc}
\usepackage[english]{babel}
\usepackage{balance}
\usepackage{cite}
\usepackage{import}
\usepackage{verbatim}

\usepackage[hidelinks]{hyperref}
\hypersetup{pdfauthor={Andreas Schwarz, Walter Kellermann},pdftitle={Coherent-to-Diffuse Power Ratio Estimation for Dereverberation}}

\usepackage{graphicx}
\usepackage{tabularx}
\usepackage{booktabs}
\usepackage{url}

\usepackage[cmex10]{amsmath}
\usepackage{amssymb}
\usepackage{array} 
\usepackage[caption=false,font=footnotesize]{subfig}
\usepackage{stfloats}

\newcounter{MYtempeqncnt}

\renewcommand\Re{\operatorname{Re}}
\renewcommand\Im{\operatorname{Im}}

\newcommand{\frameix}{{l}}

\newcommand{\dB}{{\mathrm{dB}}}
\newcommand{\CDR}{{\mathit{CDR}}}
\newcommand{\CDRjeub}{{\widehat{\CDR}_\text{Jeub}}}
\newcommand{\CDRthiergartone}{{\widehat{\CDR}_\text{Thiergart,1}}}
\newcommand{\CDRthiergarttwo}{{\widehat{\CDR}_\text{Thiergart,2}}}
\newcommand{\CDRpropone}{{\widehat{\CDR}_\text{prop1}}}
\newcommand{\CDRproptwo}{{\widehat{\CDR}_\text{prop2}}}
\newcommand{\CDRpropthree}{{\widehat{\CDR}_\text{prop3}}}
\newcommand{\CDRpropfour}{{\widehat{\CDR}_\text{prop4}}}
\newcommand{\SNR}{{\mathit{SNR}}}
\newcommand{\ELR}{{\mathit{ELR}}}
\newcommand{\MSE}{{\mathit{MSE}}}
\newcommand{\fwSegSNR}{{\mathit{fwSegSNR}}}
\newcommand{\fwSegSDR}{{\mathit{fwSegSDR}}}

\newcommand{\E}{{\mathcal{E}}}

\usepackage{tikz}

\usepackage{pgfplots}
\usepgfplotslibrary{groupplots}
\usetikzlibrary{plotmarks}
\pgfplotsset{compat=newest}
\pgfplotsset{plot coordinates/math parser=false}
\pgfplotsset{/pgf/number format/set thousands separator={}}
\newlength\figureheight
\newlength\figurewidth

\newenvironment{customlegend}[1][]{%
    \begingroup
    \csname pgfplots@init@cleared@structures\endcsname
    \pgfplotsset{#1}%
}{%
    \csname pgfplots@createlegend\endcsname
    \endgroup
}%

\def\addlegendimage{\csname pgfplots@addlegendimage\endcsname}

\begin{document}
\title{Coherent-to-Diffuse Power Ratio Estimation\\for Dereverberation}%
\author{Andreas Schwarz*, Walter Kellermann, \IEEEmembership{Fellow, IEEE}%
\thanks{A. Schwarz and W. Kellermann are with the Chair of Multimedia Communications and Signal Processing, Friedrich-Alexander-Universität Erlangen-Nürnberg, 91058 Erlangen, Germany (e-mail: schwarz@LNT.de; wk@LNT.de).}\thanks{The authors would like to thank Opticom GmbH for providing PESQ evaluation software.}\thanks{EDICS: AUD-SIRR, AUD-MAAE, AUD-SEN, AUD-ASAP}}
\maketitle

\begin{abstract}
The estimation of the time- and frequency-dependent coherent-to-diffuse power ratio (CDR) from the measured spatial coherence between two omnidirectional microphones is investigated. Known CDR estimators are formulated in a common framework, illustrated using a geometric interpretation in the complex plane, and investigated with respect to bias and robustness towards model errors. Several novel unbiased CDR estimators are proposed, and it is shown that knowledge of either the direction of arrival (DOA) of the target source or the coherence of the noise field is sufficient for unbiased CDR estimation. The validity of the model for the application of CDR estimates to dereverberation is investigated using measured and simulated impulse responses. A CDR-based dereverberation system is presented and evaluated using signal-based quality measures as well as automatic speech recognition accuracy. The results show that the proposed unbiased estimators have a practical advantage over existing estimators, and that the proposed DOA-independent estimator can be used for effective blind dereverberation.
\end{abstract}
\begin{IEEEkeywords}
Spatial Coherence, Diffuse Noise Suppression, Diffuseness, Dereverberation, Reverberation Suppression
\end{IEEEkeywords}

\section{Introduction}
\label{sec:Introduction}

\IEEEPARstart{I}{t} has been observed as early as 1969 that the measured spatial coherence between two microphones allows the discrimination between direct sound and reverberation \cite{danilenko_binaurales_1969}. A first signal enhancement algorithm based on this observation was proposed by Allen et al. in 1977 \cite{allen_multimicrophone_1977}, where the magnitude of the coherence is estimated in the Short-Time Fourier Transform (STFT) domain and used as a gain for reverberation suppression. Other heuristic methods for noise reduction and dereverberation using coherence estimates have since been proposed \cite{bloom_evaluation_1982, zelinski_microphone_1988, le_bouquin_using_1992, le_bouquin-jeannes_enhancement_1997,westermann_binaural_2013}. %
Related methods have also been investigated for noise suppression in connection with beamforming, and postfilters which are statistically optimal under certain conditions have been proposed for the suppression of uncorrelated \cite{simmer_post-filtering_2001} and diffuse \cite{mccowan_microphone_2003} noise.

More recently, explicit estimators for the ratio between direct and diffuse signal components, termed the coherent-to-diffuse power ratio (CDR), from short-time coherence estimates have been formulated \cite{jeub_blind_2011,thiergart_signal--reverberant_2012}, based on the same assumptions as the earlier optimum postfilter derivations \cite{mccowan_microphone_2003}. Also, results have since been generalized from omnidirectional microphones to other microphone directivities \cite{thiergart_diffuseness_2011, thiergart_spatial_2012} and spherical microphone arrays \cite{jarrett_coherence-based_2012}. %
While these estimates can be used for the formulation of postfilters for signal enhancement \cite{schwarz_unbiased_2014}, which is the main application considered in this contribution, short-time CDR estimates (or the equivalent ``diffuseness'' measure) also have applications in parametric coding of spatial audio signals \cite{pulkki_spatial_2007} and the extraction of spatial features for automatic speech recognition (ASR) \cite{schwarz_spatial_2014}.

In this contribution, the estimation of the CDR from the measured coherence between two omnidirectional microphones, and the application of the CDR estimates to dereverberation, is investigated. First, the signal model for the recording of a noisy or reverberant signal with two omnidirectional microphones is described, the relationship between signal and noise coherence models and the coherence of the mixed signal is given, and coherence models for the application to dereverberation are discussed. Then, several known CDR estimators are formulated in a common framework, illustrated using a geometric interpretation in the complex plane, and improved unbiased estimators are proposed. It is shown that knowledge of either the target signal direction or the noise coherence is sufficient for an unbiased CDR estimation, and estimators are proposed for the cases of unknown target signal direction and unknown noise coherence. Finally, the CDR estimators are applied in a postfilter for reverberation suppression and evaluated by processing reverberant speech and comparing ASR recognition accuracy as well as various signal quality measures.
This paper builds on results published in a recent conference paper by the same authors, in which the novel estimators were initially proposed \cite{schwarz_unbiased_2014}.

\section{Signal Model}
\label{sec:signal-model}

We consider the recording of a reverberant or noisy speech signal by two omnidirectional microphones with a spacing $d$, located in the same horizontal plane.
The signal $x_i(t)$ of the $i$-th microphone is composed of a desired signal component $s_i(t)$ and an undesired component $n_i(t)$ consisting of noise and/or late reverberation, i.e.,
\begin{flalign}
x_i(t) = s_i(t) + n_i(t),~i=1,2.
\end{flalign}
The microphone, desired and noise signals are represented in the time-frequency (STFT) domain by the corresponding uppercase letters, i.e., $X_i(\frameix,f)$, $S_i(\frameix,f)$ and $N_i(\frameix,f)$, respectively, with the discrete-time frame index $\frameix$ and continuous frequency $f$, and are assumed to be short-time stationary. Using the representation in the STFT domain, the short-time auto- and cross-power spectra between two signals $u(t)$ and $v(t)$ are defined as
\begin{flalign}
\Phi_{u v}(\frameix,f)=\E\{U(\frameix,f) V^*(\frameix,f)\},
\end{flalign}
where $\E$ is the expectation operator.
It is assumed that the auto-power spectra of the signal components are the same at both microphones, i.e.,
\begin{flalign}
\Phi_{s_1 s_1}(\frameix,f) &= \Phi_{s_2 s_2}(\frameix,f) =\Phi_{s}(\frameix,f),\\
\Phi_{n_1 n_1}(\frameix,f) &= \Phi_{n_2 n_2}(\frameix,f) =\Phi_{n}(\frameix,f).
\end{flalign}
Note that this assumption is generally appropriate for a plane wave as desired signal as well as for noise and late reverberation, but may in practice be impacted by the presence of early reflections causing destructive or constructive interference.
The time- and frequency-dependent signal-to-noise ratio (SNR) of the microphone signals can be defined as
\begin{flalign}
\SNR(\frameix,f)=\frac{\Phi_{s}(\frameix,f)}{\Phi_{n}(\frameix,f)}.
\end{flalign}
The complex spatial coherence functions of the desired signal and noise components are given by
\begin{flalign}
\Gamma_s(f)=\frac{\Phi_{s_1 s_2}(\frameix,f)}{\Phi_{s}(\frameix,f)},
\Gamma_n(f)=\frac{\Phi_{n_1 n_2}(\frameix,f)}{\Phi_{n}(\frameix,f)},
\end{flalign}
respectively, and are assumed to be time-invariant, i.e., dependent only on the spatial characteristics of the signal components. It is furthermore assumed that signal and noise are mutually orthogonal, such that
\begin{flalign}
\Phi_x(\frameix,f) = \Phi_s(\frameix,f) + \Phi_n(\frameix,f).
\end{flalign}
The complex spatial coherence of the mixed sound field can then be written as a function of the SNR and the signal and noise coherence functions:
\begin{flalign}
\Gamma_{x}(\frameix,f) = \frac{ \SNR(\frameix,f) \Gamma_{s}(f) + \Gamma_{n}(f) }{ \SNR(\frameix,f) + 1 }.
\label{eq:Gamma_x}
\end{flalign}
This relationship is valid for any signal and noise coherence function. For the special case of a fully coherent desired signal component and diffuse noise, the term CDR or direct-to-diffuse ratio (DDR) is often used for the SNR. We will adopt the term CDR in the following.
(\ref{eq:Gamma_x}) can be rewritten as a parametric line equation in the complex plane, highlighting that $\Gamma_x$ lies on a straight line connecting $\Gamma_n$ and $\Gamma_s$:
\begin{flalign}
\Gamma_{x}(\frameix,f) = \Gamma_{s}(f) + \frac{1}{\CDR(\frameix,f)+1} (\Gamma_{n}(f) - \Gamma_{s}(f)).
\label{eq:Gamma_x_line}
\end{flalign}
Note that the line parameter $D(\frameix,f)=[\CDR(\frameix,f)+1]^{-1}$ is equivalent to the \emph{diffuseness} defined in \cite{del_galdo_diffuse_2012}. %

\section{Coherence Models for Dereverberation}

The desired and noise or reverberation components of the microphone signals are characterized by time-invariant coherence functions $\Gamma_s(f)$ and $\Gamma_n(f)$, respectively. In the following, suitable models for these spatial coherence functions are discussed for the application to dereverberation.

\subsection{Desired Signal}
The desired signal component is modeled as a plane wave with the direction of arrival (DOA) $\theta$ with respect to the microphone axis, where $\theta=0\,^\circ$ corresponds to broadside direction. The corresponding time-invariant coherence function is given by
\begin{flalign}
\label{eq:coh-direct}
\Gamma_s(f) &= \frac{\Phi_{s_1 s_2}(\frameix,f)}{\Phi_s(\frameix,f)} = e^{j k d \sin(\theta)} = e^{j 2 \pi f \Delta t},
\end{flalign}
with the time difference of arrival (TDOA) $\Delta t = d \sin(\theta)/c$, the wavenumber $k=2 \pi f/c$ and the speed of sound $c$.
This coherence function always has a magnitude of one, and is equal to one for $\Delta t=0$.

\subsection{Reverberation as Isotropic Sound Field}
\label{sec:reverberation-coherence}
In array signal processing, environmental noise is often modeled by the superposition of an infinite number of uncorrelated, spatially distributed noise sources. In applications like underwater acoustics or radio communication, this model is motivated by the presence of many independent noise and interfering sources around the receiver \cite{cron_spatial-correlation_1962}. The most common assumption for the spatial distribution is a sphere centered around the receiver, which corresponds to what is known as a \emph{diffuse} or \emph{spherically isotropic} noise field. The spatial coherence function between two omnidirectional sensors in a diffuse noise field is real-valued and given by
\begin{flalign}
\label{eq:coh-diffuse}
\Gamma_\text{diffuse}(f) = \frac{\sin(k d)}{k d} = \frac{\sin(2 \pi f d/c)}{2 \pi f d/c}.
\end{flalign}
While diffusivity of the noise field is easily motivated in the aforementioned scenarios, a few more considerations are necessary for the modeling of a reverberation component originating from a single excitation signal. Since acoustic transmission within a room is generally assumed to be linear and time-invariant, a reverberant signal can be modeled by the convolution of a source signal with a time-invariant room impulse response (RIR) \cite{kuttruff_room_2000}. The reverberant signals recorded at two points in space, i.e., by two microphones, are therefore linearly related, and the theoretical coherence function between these two signals is equal to one. However, when limited observation windows are considered, and the excitation signal has a limited temporal correlation, reflections with different delays can be approximated as uncorrelated sources. This uncorrelated scattering assumption is widely used in mobile radio communications \cite{steele_mobile_1999} and underwater acoustics \cite{kilfoyle_state_2000}, and is useful in room acoustics as well, where it has been observed that the sound field in a reverberant room appears as an approximately diffuse sound field \cite{cook_measurement_1955,jacobsen_coherence_2000}. The plausibility of the diffuseness assumption for reverberation can be visualized using the image source model \cite{allen_image_1979}: for higher reflection orders, the angular distribution of the image sources becomes increasingly isotropic. Furthermore, given a limited observation window length, the delayed reflected versions of the source signal are increasingly decorrelated with increasing reflection orders. Based on this idea, we can predict a number of factors which contribute to how well the model of diffuseness is fulfilled: a large room contributes to the uncorrelatedness of the image sources, due to larger relative delays between reflections; highly reflective surfaces contribute to the presence of many image sources with similar power, since the power contributed by reflections decays more slowly with the reflection order;
and low temporal correlation of the source signal contributes to low correlation between the delayed reflections. Some of these effects are illustrated in Section~\ref{sec:evaluation-reverberation-coherence} using measured and simulated RIRs.

In real rooms, effects like diffraction, diffuse reflection \cite{kuttruff_room_2000}, and potentially time-variant effects \cite{elko_room_2003} may further contribute to the randomization of delays and incidence angles of reflections and therefore increase the diffuseness of the reverberation sound field. However, as shown later, the image source model is sufficient to explain a wide range of practical effects which affect the reverberation coherence.

While the diffuse sound field model is the most common in room acoustics and signal enhancement, it has been observed that reverberant noise in rooms with highly absorbing floors and ceilings can be modeled more accurately by noise sources distributed in the horizontal plane, i.e., by a 2D isotropic (cylindrically isotropic) noise field, as opposed to a diffuse (spherically isotropic) noise field \cite{elko_superdirectional_2000}. This noise field model consists of uncorrelated noise sources located on a circle around and in the same plane as the microphones (typically the horizontal plane), and is motivated by the rapid decay of all vertically propagating sound components due to the strong absorption at the floor and/or ceiling. The corresponding spatial coherence function for two omnidirectional microphones located in the same plane as the noise sources is the zeroth-order Bessel function of the first kind \cite{cook_measurement_1955, elko_spatial_2001}:
\begin{flalign}
\Gamma_\text{2D-iso}(f) = J_0(k d) = J_0(2 \pi f d/c).
\end{flalign}

Note that, both in the case of diffuse and 2D-isotropic noise fields, the coherence function is real-valued, since the spatial distribution of the sources is symmetric with respect to the microphone array axis.

In Section~\ref{sec:evaluation-reverberation-coherence}, the effects of room geometry and surface reflectivity on the coherence of the reverberation component are evaluated using RIRs generated with the image source method, and RIRs that were measured in different rooms.

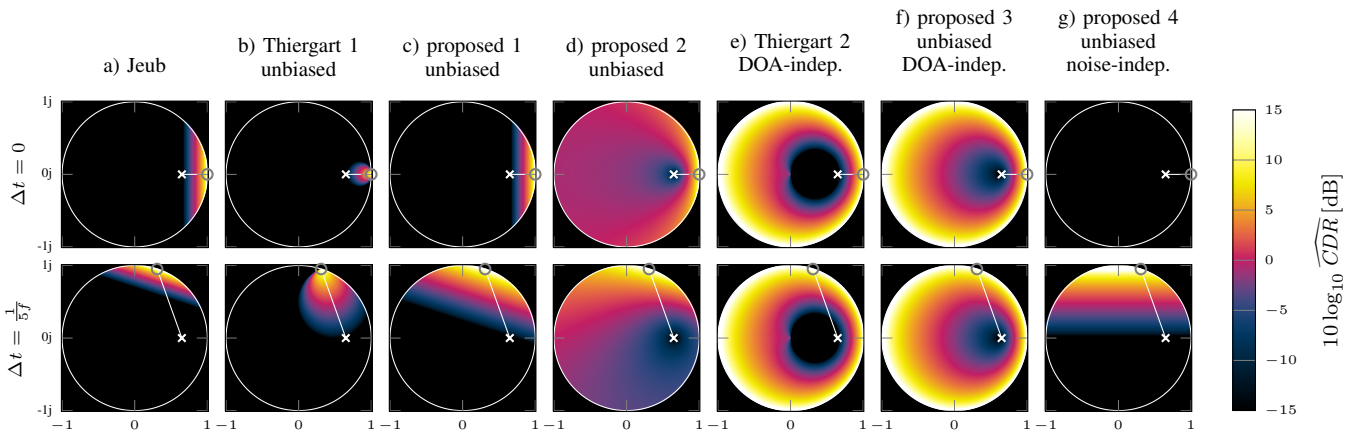
\begin{figure*}[b]
    \centering
        \setlength\figureheight{1.95cm}
	\setlength\figurewidth{1.95cm}
	\pgfplotsset{
	tick label style={font=\tiny},
	label style={font=\footnotesize},
	legend style={font=\footnotesize},
	title style={align=center,font=\footnotesize,text height=1em},
	}
\begin{tikzpicture}
    \node(scope1){
\begin{tikzpicture}
\hspace{-3mm}
\matrix[row sep=-2mm, column sep=-2mm,name=plotmatrix]{
\node[rotate=90,font=\scriptsize,align=center,inner ysep=5pt] at (0,0.5*\figureheight) {$\Delta t = 0$}; &

\begin{axis}[%
width=\figurewidth,
height=\figureheight,
axis on top,
scale only axis,
xmin=-1.01,
xmax=1.01,
xtick={-1,0,1},
xticklabels={\empty},
ymin=-1.01,
ymax=1.01,
ytick={-1,0,1},
yticklabels={{-1j},{0j},{1j}},
title={a) Jeub}
]
\addplot [forget plot] graphics [xmin=-1.01202404809619,xmax=1.01202404809619,ymin=-1.01202404809619,ymax=1.01202404809619] {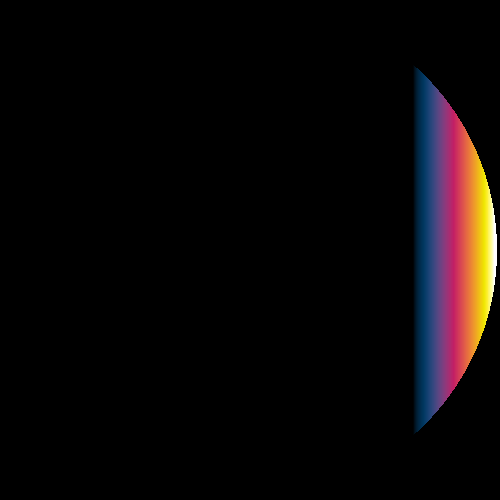};
\addplot [color=white,solid,forget plot]
  table[row sep=crcr]{figures/heatmaps-1-1-1.tsv};
\addplot [color=gray,line width=0.8pt,only marks,mark=o,mark options={solid},forget plot]
  table[row sep=crcr]{figures/heatmaps-1-1-2.tsv};
\addplot [color=white,line width=0.8pt,only marks,mark=x,mark options={solid},forget plot]
  table[row sep=crcr]{figures/heatmaps-1-1-3.tsv};
\addplot [color=white,solid,forget plot]
  table[row sep=crcr]{figures/heatmaps-1-1-4.tsv};
\end{axis}
 &

\begin{axis}[%
width=\figurewidth,
height=\figureheight,
axis on top,
scale only axis,
xmin=-1.01,
xmax=1.01,
xtick={-1,0,1},
xticklabels={\empty},
ymin=-1.01,
ymax=1.01,
ytick={-1,0,1},
yticklabels={\empty},
title={b) Thiergart 1\\unbiased}
]
\addplot [forget plot] graphics [xmin=-1.01202404809619,xmax=1.01202404809619,ymin=-1.01202404809619,ymax=1.01202404809619] {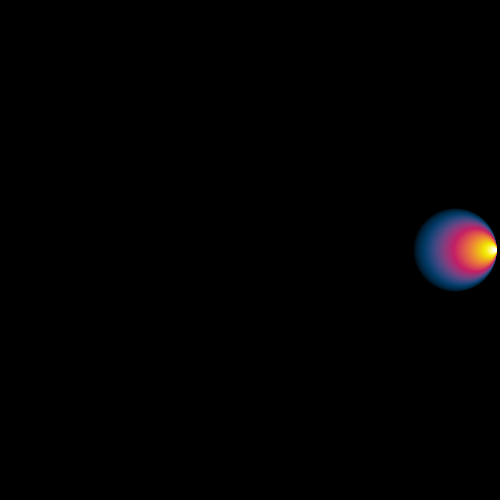};
\addplot [color=white,solid,forget plot]
  table[row sep=crcr]{figures/heatmaps-1-2-1.tsv};
\addplot [color=gray,line width=0.8pt,only marks,mark=o,mark options={solid},forget plot]
  table[row sep=crcr]{figures/heatmaps-1-2-2.tsv};
\addplot [color=white,line width=0.8pt,only marks,mark=x,mark options={solid},forget plot]
  table[row sep=crcr]{figures/heatmaps-1-2-3.tsv};
\addplot [color=white,solid,forget plot]
  table[row sep=crcr]{figures/heatmaps-1-2-4.tsv};
\end{axis}
 &

\begin{axis}[%
width=\figurewidth,
height=\figureheight,
axis on top,
scale only axis,
xmin=-1.01,
xmax=1.01,
xtick={-1,0,1},
xticklabels={\empty},
ymin=-1.01,
ymax=1.01,
ytick={-1,0,1},
yticklabels={\empty},
title={c) proposed 1\\unbiased}
]
\addplot [forget plot] graphics [xmin=-1.01202404809619,xmax=1.01202404809619,ymin=-1.01202404809619,ymax=1.01202404809619] {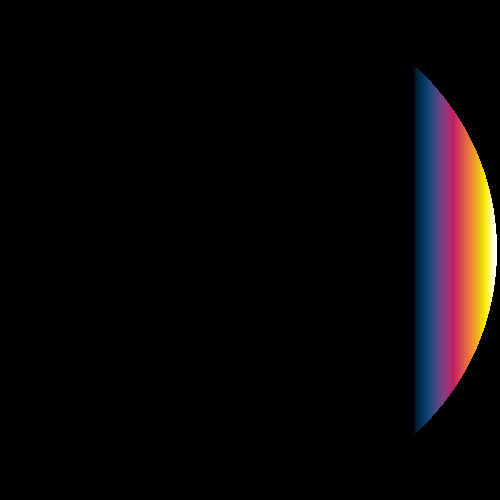};
\addplot [color=white,solid,forget plot]
  table[row sep=crcr]{figures/heatmaps-1-3-1.tsv};
\addplot [color=gray,line width=0.8pt,only marks,mark=o,mark options={solid},forget plot]
  table[row sep=crcr]{figures/heatmaps-1-3-2.tsv};
\addplot [color=white,line width=0.8pt,only marks,mark=x,mark options={solid},forget plot]
  table[row sep=crcr]{figures/heatmaps-1-3-3.tsv};
\addplot [color=white,solid,forget plot]
  table[row sep=crcr]{figures/heatmaps-1-3-4.tsv};
\end{axis}
 &

\begin{axis}[%
width=\figurewidth,
height=\figureheight,
axis on top,
scale only axis,
xmin=-1.01,
xmax=1.01,
xtick={-1,0,1},
xticklabels={\empty},
ymin=-1.01,
ymax=1.01,
ytick={-1,0,1},
yticklabels={\empty},
title={d) proposed 2\\unbiased}
]
\addplot [forget plot] graphics [xmin=-1.01202404809619,xmax=1.01202404809619,ymin=-1.01202404809619,ymax=1.01202404809619] {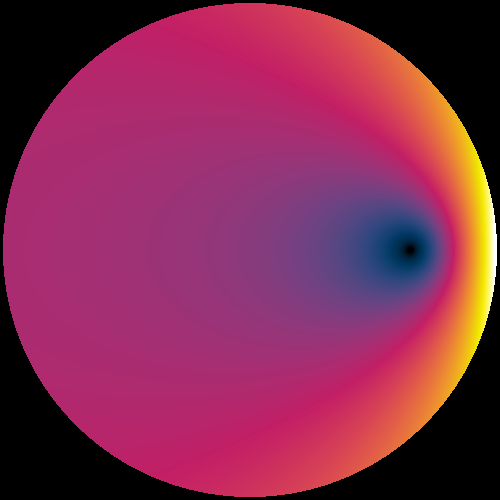};
\addplot [color=white,solid,forget plot]
  table[row sep=crcr]{figures/heatmaps-1-4-1.tsv};
\addplot [color=gray,line width=0.8pt,only marks,mark=o,mark options={solid},forget plot]
  table[row sep=crcr]{figures/heatmaps-1-4-2.tsv};
\addplot [color=white,line width=0.8pt,only marks,mark=x,mark options={solid},forget plot]
  table[row sep=crcr]{figures/heatmaps-1-4-3.tsv};
\addplot [color=white,solid,forget plot]
  table[row sep=crcr]{figures/heatmaps-1-4-4.tsv};
\end{axis}
 &

\begin{axis}[%
width=\figurewidth,
height=\figureheight,
axis on top,
scale only axis,
xmin=-1.01,
xmax=1.01,
xtick={-1,0,1},
xticklabels={\empty},
ymin=-1.01,
ymax=1.01,
ytick={-1,0,1},
yticklabels={\empty},
title={e) Thiergart 2\\DOA-indep.}
]
\addplot [forget plot] graphics [xmin=-1.01202404809619,xmax=1.01202404809619,ymin=-1.01202404809619,ymax=1.01202404809619] {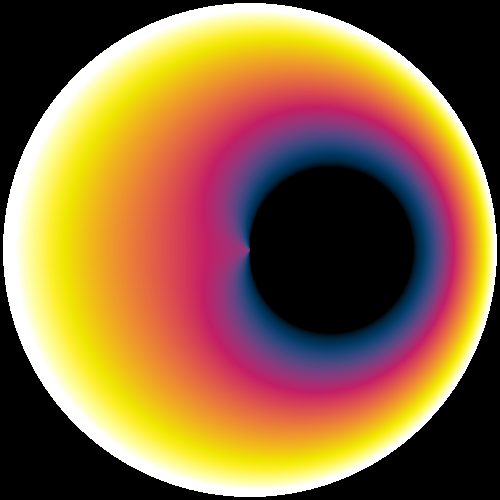};
\addplot [color=white,solid,forget plot]
  table[row sep=crcr]{figures/heatmaps-1-5-1.tsv};
\addplot [color=gray,line width=0.8pt,only marks,mark=o,mark options={solid},forget plot]
  table[row sep=crcr]{figures/heatmaps-1-5-2.tsv};
\addplot [color=white,line width=0.8pt,only marks,mark=x,mark options={solid},forget plot]
  table[row sep=crcr]{figures/heatmaps-1-5-3.tsv};
\addplot [color=white,solid,forget plot]
  table[row sep=crcr]{figures/heatmaps-1-5-4.tsv};
\end{axis}
 &

\begin{axis}[%
width=\figurewidth,
height=\figureheight,
axis on top,
scale only axis,
xmin=-1.01,
xmax=1.01,
xtick={-1,0,1},
xticklabels={\empty},
ymin=-1.01,
ymax=1.01,
ytick={-1,0,1},
yticklabels={\empty},
title={f) proposed 3\\unbiased\\DOA-indep.}
]
\addplot [forget plot] graphics [xmin=-1.01202404809619,xmax=1.01202404809619,ymin=-1.01202404809619,ymax=1.01202404809619] {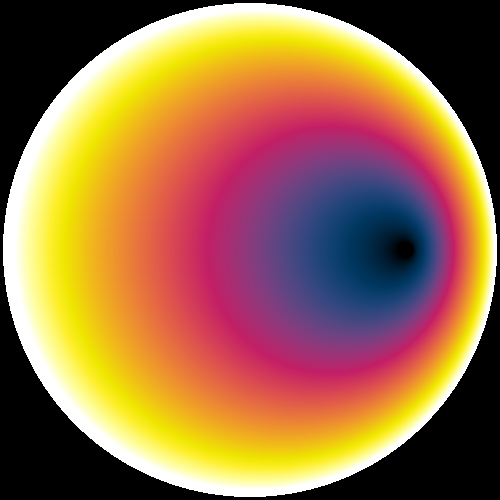};
\addplot [color=white,solid,forget plot]
  table[row sep=crcr]{figures/heatmaps-1-6-1.tsv};
\addplot [color=gray,line width=0.8pt,only marks,mark=o,mark options={solid},forget plot]
  table[row sep=crcr]{figures/heatmaps-1-6-2.tsv};
\addplot [color=white,line width=0.8pt,only marks,mark=x,mark options={solid},forget plot]
  table[row sep=crcr]{figures/heatmaps-1-6-3.tsv};
\addplot [color=white,solid,forget plot]
  table[row sep=crcr]{figures/heatmaps-1-6-4.tsv};
\end{axis}
 &

\begin{axis}[%
width=\figurewidth,
height=\figureheight,
axis on top,
scale only axis,
xmin=-1.01,
xmax=1.01,
xtick={-1,0,1},
xticklabels={\empty},
ymin=-1.01,
ymax=1.01,
ytick={-1,0,1},
yticklabels={\empty},
title={g) proposed 4\\unbiased\\noise-indep.}
]
\addplot [forget plot] graphics [xmin=-1.01202404809619,xmax=1.01202404809619,ymin=-1.01202404809619,ymax=1.01202404809619] {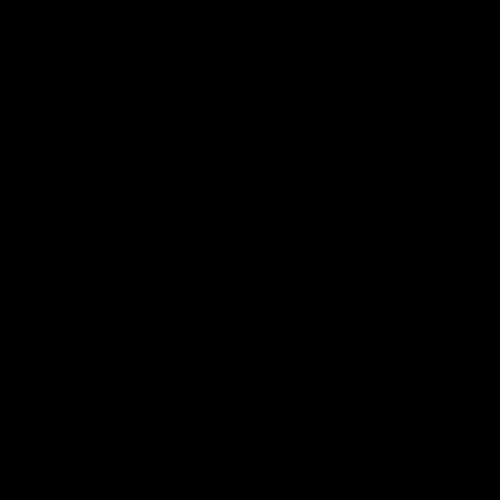};
\addplot [color=white,solid,forget plot]
  table[row sep=crcr]{figures/heatmaps-1-7-1.tsv};
\addplot [color=gray,line width=0.8pt,only marks,mark=o,mark options={solid},forget plot]
  table[row sep=crcr]{figures/heatmaps-1-7-2.tsv};
\addplot [color=white,line width=0.8pt,only marks,mark=x,mark options={solid},forget plot]
  table[row sep=crcr]{figures/heatmaps-1-7-3.tsv};
\addplot [color=white,solid,forget plot]
  table[row sep=crcr]{figures/heatmaps-1-7-4.tsv};
\end{axis}
 \\
\node[rotate=90,font=\scriptsize,align=center,inner ysep=5pt] at (0,0.5*\figureheight) {$\Delta t = \frac{1}{5 f}$}; &

\begin{axis}[%
width=\figurewidth,
height=\figureheight,
axis on top,
scale only axis,
xmin=-1.01,
xmax=1.01,
xtick={-1,0,1},
ymin=-1.01,
ymax=1.01,
ytick={-1,0,1},
yticklabels={{-1j},{0j},{1j}}
]
\addplot [forget plot] graphics [xmin=-1.01202404809619,xmax=1.01202404809619,ymin=-1.01202404809619,ymax=1.01202404809619] {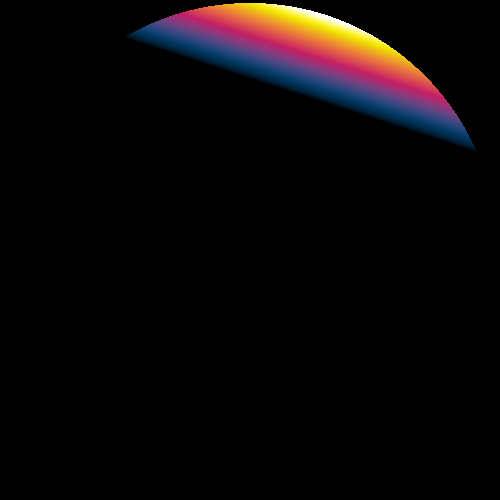};
\addplot [color=white,solid,forget plot]
  table[row sep=crcr]{figures/heatmaps-2-1-1.tsv};
\addplot [color=gray,line width=0.8pt,only marks,mark=o,mark options={solid},forget plot]
  table[row sep=crcr]{figures/heatmaps-2-1-2.tsv};
\addplot [color=white,line width=0.8pt,only marks,mark=x,mark options={solid},forget plot]
  table[row sep=crcr]{figures/heatmaps-2-1-3.tsv};
\addplot [color=white,solid,forget plot]
  table[row sep=crcr]{figures/heatmaps-2-1-4.tsv};
\end{axis}
 &

\begin{axis}[%
width=\figurewidth,
height=\figureheight,
axis on top,
scale only axis,
xmin=-1.01,
xmax=1.01,
xtick={-1,0,1},
ymin=-1.01,
ymax=1.01,
ytick={-1,0,1},
yticklabels={\empty}
]
\addplot [forget plot] graphics [xmin=-1.01202404809619,xmax=1.01202404809619,ymin=-1.01202404809619,ymax=1.01202404809619] {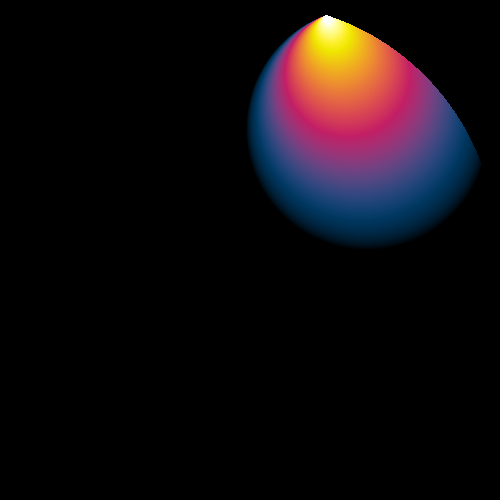};
\addplot [color=white,solid,forget plot]
  table[row sep=crcr]{figures/heatmaps-2-2-1.tsv};
\addplot [color=gray,line width=0.8pt,only marks,mark=o,mark options={solid},forget plot]
  table[row sep=crcr]{figures/heatmaps-2-2-2.tsv};
\addplot [color=white,line width=0.8pt,only marks,mark=x,mark options={solid},forget plot]
  table[row sep=crcr]{figures/heatmaps-2-2-3.tsv};
\addplot [color=white,solid,forget plot]
  table[row sep=crcr]{figures/heatmaps-2-2-4.tsv};
\end{axis}
 &

\begin{axis}[%
width=\figurewidth,
height=\figureheight,
axis on top,
scale only axis,
xmin=-1.01,
xmax=1.01,
xtick={-1,0,1},
ymin=-1.01,
ymax=1.01,
ytick={-1,0,1},
yticklabels={\empty}
]
\addplot [forget plot] graphics [xmin=-1.01202404809619,xmax=1.01202404809619,ymin=-1.01202404809619,ymax=1.01202404809619] {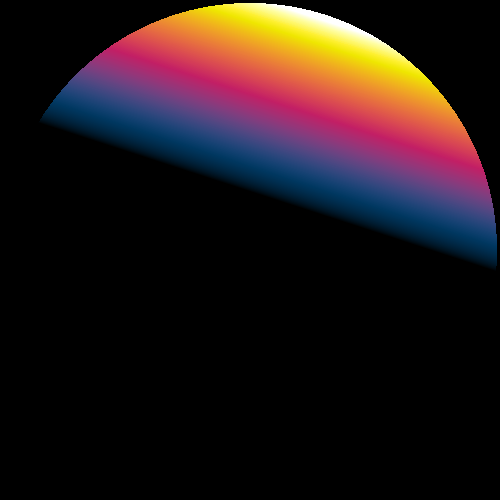};
\addplot [color=white,solid,forget plot]
  table[row sep=crcr]{figures/heatmaps-2-3-1.tsv};
\addplot [color=gray,line width=0.8pt,only marks,mark=o,mark options={solid},forget plot]
  table[row sep=crcr]{figures/heatmaps-2-3-2.tsv};
\addplot [color=white,line width=0.8pt,only marks,mark=x,mark options={solid},forget plot]
  table[row sep=crcr]{figures/heatmaps-2-3-3.tsv};
\addplot [color=white,solid,forget plot]
  table[row sep=crcr]{figures/heatmaps-2-3-4.tsv};
\end{axis}
 &

\begin{axis}[%
width=\figurewidth,
height=\figureheight,
axis on top,
scale only axis,
xmin=-1.01,
xmax=1.01,
xtick={-1,0,1},
ymin=-1.01,
ymax=1.01,
ytick={-1,0,1},
yticklabels={\empty}
]
\addplot [forget plot] graphics [xmin=-1.01202404809619,xmax=1.01202404809619,ymin=-1.01202404809619,ymax=1.01202404809619] {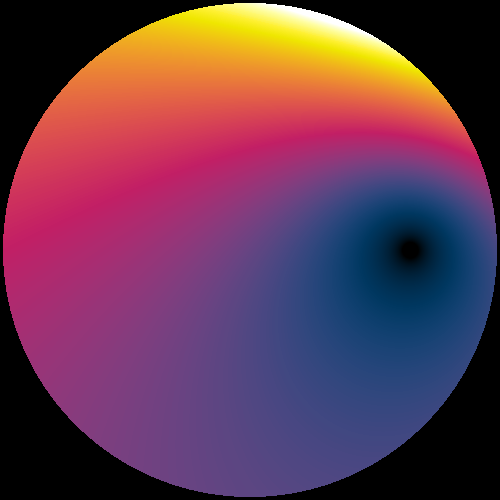};
\addplot [color=white,solid,forget plot]
  table[row sep=crcr]{figures/heatmaps-2-4-1.tsv};
\addplot [color=gray,line width=0.8pt,only marks,mark=o,mark options={solid},forget plot]
  table[row sep=crcr]{figures/heatmaps-2-4-2.tsv};
\addplot [color=white,line width=0.8pt,only marks,mark=x,mark options={solid},forget plot]
  table[row sep=crcr]{figures/heatmaps-2-4-3.tsv};
\addplot [color=white,solid,forget plot]
  table[row sep=crcr]{figures/heatmaps-2-4-4.tsv};
\end{axis}
 &

\begin{axis}[%
width=\figurewidth,
height=\figureheight,
axis on top,
scale only axis,
xmin=-1.01,
xmax=1.01,
xtick={-1,0,1},
ymin=-1.01,
ymax=1.01,
ytick={-1,0,1},
yticklabels={\empty}
]
\addplot [forget plot] graphics [xmin=-1.01202404809619,xmax=1.01202404809619,ymin=-1.01202404809619,ymax=1.01202404809619] {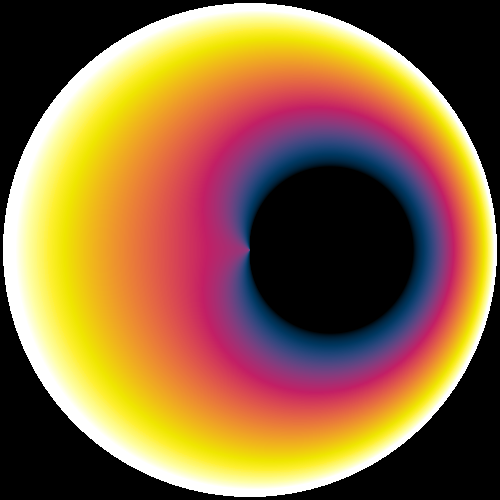};
\addplot [color=white,solid,forget plot]
  table[row sep=crcr]{figures/heatmaps-2-5-1.tsv};
\addplot [color=gray,line width=0.8pt,only marks,mark=o,mark options={solid},forget plot]
  table[row sep=crcr]{figures/heatmaps-2-5-2.tsv};
\addplot [color=white,line width=0.8pt,only marks,mark=x,mark options={solid},forget plot]
  table[row sep=crcr]{figures/heatmaps-2-5-3.tsv};
\addplot [color=white,solid,forget plot]
  table[row sep=crcr]{figures/heatmaps-2-5-4.tsv};
\end{axis}
 &

\begin{axis}[%
width=\figurewidth,
height=\figureheight,
axis on top,
scale only axis,
xmin=-1.01,
xmax=1.01,
xtick={-1,0,1},
ymin=-1.01,
ymax=1.01,
ytick={-1,0,1},
yticklabels={\empty}
]
\addplot [forget plot] graphics [xmin=-1.01202404809619,xmax=1.01202404809619,ymin=-1.01202404809619,ymax=1.01202404809619] {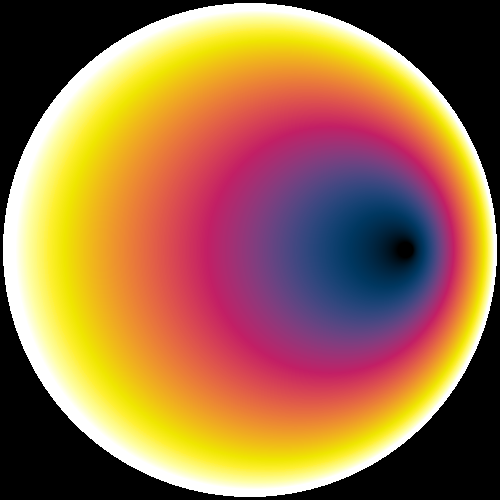};
\addplot [color=white,solid,forget plot]
  table[row sep=crcr]{figures/heatmaps-2-6-1.tsv};
\addplot [color=gray,line width=0.8pt,only marks,mark=o,mark options={solid},forget plot]
  table[row sep=crcr]{figures/heatmaps-2-6-2.tsv};
\addplot [color=white,line width=0.8pt,only marks,mark=x,mark options={solid},forget plot]
  table[row sep=crcr]{figures/heatmaps-2-6-3.tsv};
\addplot [color=white,solid,forget plot]
  table[row sep=crcr]{figures/heatmaps-2-6-4.tsv};
\end{axis}
 &

\begin{axis}[%
width=\figurewidth,
height=\figureheight,
axis on top,
scale only axis,
xmin=-1.01,
xmax=1.01,
xtick={-1,0,1},
ymin=-1.01,
ymax=1.01,
ytick={-1,0,1},
yticklabels={\empty}
]
\addplot [forget plot] graphics [xmin=-1.01202404809619,xmax=1.01202404809619,ymin=-1.01202404809619,ymax=1.01202404809619] {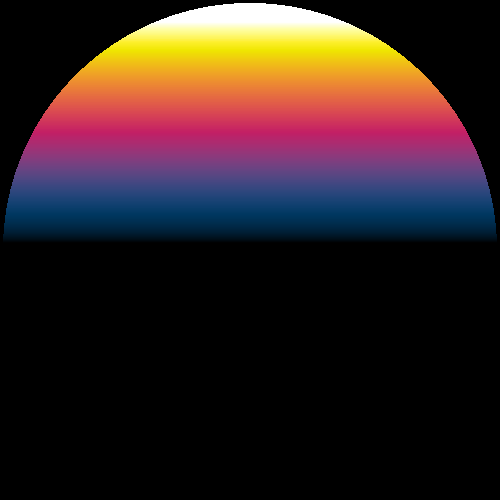};
\addplot [color=white,solid,forget plot]
  table[row sep=crcr]{figures/heatmaps-2-7-1.tsv};
\addplot [color=gray,line width=0.8pt,only marks,mark=o,mark options={solid},forget plot]
  table[row sep=crcr]{figures/heatmaps-2-7-2.tsv};
\addplot [color=white,line width=0.8pt,only marks,mark=x,mark options={solid},forget plot]
  table[row sep=crcr]{figures/heatmaps-2-7-3.tsv};
\addplot [color=white,solid,forget plot]
  table[row sep=crcr]{figures/heatmaps-2-7-4.tsv};
\end{axis}
 \\
};
\end{tikzpicture}
};
\node[at={($(scope1.east)+(-6mm,-3mm)$)},anchor=west] (scope2){
\begin{tikzpicture}
	\begin{axis}[%
	anchor=west,
	scale only axis,
	hide axis,
	width=0pt,
	height=2cm,
	colormap={mymap}{[1pt] rgb(0pt)=(0,0,0); rgb(63pt)=(0,0.21396,0.370941); rgb(64pt)=(0.00114687,0.217332,0.37684); rgb(65pt)=(0.00589468,0.220609,0.382747); rgb(66pt)=(0.0109031,0.223854,0.388609); rgb(67pt)=(0.0161776,0.227063,0.394421); rgb(68pt)=(0.0217231,0.230233,0.400178); rgb(69pt)=(0.0275446,0.233359,0.405874); rgb(70pt)=(0.0336465,0.236438,0.411501); rgb(71pt)=(0.0400331,0.239465,0.417055); rgb(72pt)=(0.0467084,0.242436,0.422528); rgb(73pt)=(0.0536758,0.245346,0.427914); rgb(74pt)=(0.0609385,0.24819,0.433205); rgb(75pt)=(0.0684995,0.250963,0.438394); rgb(76pt)=(0.0763611,0.25366,0.443474); rgb(77pt)=(0.0845254,0.256274,0.448436); rgb(78pt)=(0.092994,0.258799,0.453274); rgb(79pt)=(0.101768,0.261228,0.457979); rgb(80pt)=(0.110849,0.263556,0.462543); rgb(81pt)=(0.120237,0.265773,0.466959); rgb(82pt)=(0.129932,0.267872,0.471217); rgb(83pt)=(0.139934,0.269845,0.475311); rgb(84pt)=(0.150243,0.271683,0.479232); rgb(85pt)=(0.160856,0.273376,0.482971); rgb(86pt)=(0.171774,0.274914,0.486521); rgb(87pt)=(0.182993,0.276287,0.489873); rgb(88pt)=(0.194512,0.277483,0.49302); rgb(89pt)=(0.206327,0.278491,0.495953); rgb(90pt)=(0.218436,0.279299,0.498663); rgb(91pt)=(0.230834,0.279894,0.501143); rgb(92pt)=(0.243517,0.280265,0.503385); rgb(93pt)=(0.256478,0.280399,0.505379); rgb(94pt)=(0.269713,0.280283,0.507117); rgb(95pt)=(0.283213,0.279904,0.508592); rgb(96pt)=(0.29697,0.279252,0.509793); rgb(97pt)=(0.310974,0.278314,0.510713); rgb(98pt)=(0.325215,0.277082,0.511342); rgb(99pt)=(0.339681,0.275545,0.511674); rgb(100pt)=(0.354358,0.273695,0.511698); rgb(101pt)=(0.369231,0.271528,0.511408); rgb(102pt)=(0.384283,0.269038,0.510795); rgb(103pt)=(0.399496,0.266223,0.509854); rgb(104pt)=(0.414849,0.263082,0.50858); rgb(105pt)=(0.430322,0.259618,0.506967); rgb(106pt)=(0.445892,0.255833,0.505013); rgb(107pt)=(0.461535,0.251735,0.502717); rgb(108pt)=(0.477226,0.247332,0.50008); rgb(109pt)=(0.492939,0.242633,0.497104); rgb(110pt)=(0.508648,0.23765,0.493796); rgb(111pt)=(0.524328,0.232398,0.490161); rgb(112pt)=(0.539953,0.22689,0.486209); rgb(113pt)=(0.555497,0.221143,0.481952); rgb(114pt)=(0.570938,0.215173,0.477403); rgb(115pt)=(0.586251,0.208998,0.472575); rgb(116pt)=(0.601418,0.202633,0.467486); rgb(117pt)=(0.616418,0.196096,0.462152); rgb(118pt)=(0.631235,0.189403,0.45659); rgb(119pt)=(0.645854,0.18257,0.450818); rgb(120pt)=(0.660264,0.175611,0.444853); rgb(121pt)=(0.674454,0.168542,0.438713); rgb(122pt)=(0.688416,0.161373,0.432415); rgb(123pt)=(0.702144,0.154118,0.425974); rgb(124pt)=(0.715634,0.146787,0.419406); rgb(125pt)=(0.728885,0.139389,0.412723); rgb(126pt)=(0.741894,0.131933,0.40594); rgb(127pt)=(0.754663,0.124426,0.399068); rgb(128pt)=(0.765169,0.126991,0.392927); rgb(129pt)=(0.77342,0.13968,0.387501); rgb(130pt)=(0.781443,0.152391,0.381981); rgb(131pt)=(0.789236,0.165123,0.376376); rgb(132pt)=(0.7968,0.17787,0.370695); rgb(133pt)=(0.804137,0.19063,0.364946); rgb(134pt)=(0.811248,0.203401,0.359136); rgb(135pt)=(0.818135,0.216181,0.353271); rgb(136pt)=(0.824799,0.228969,0.347357); rgb(137pt)=(0.831243,0.241763,0.341399); rgb(138pt)=(0.83747,0.254562,0.335401); rgb(139pt)=(0.84348,0.267365,0.329369); rgb(140pt)=(0.849278,0.280172,0.323304); rgb(141pt)=(0.854866,0.292981,0.317211); rgb(142pt)=(0.860246,0.305792,0.311092); rgb(143pt)=(0.865421,0.318604,0.304949); rgb(144pt)=(0.870394,0.331417,0.298784); rgb(145pt)=(0.875167,0.344229,0.2926); rgb(146pt)=(0.879743,0.35704,0.286397); rgb(147pt)=(0.884125,0.369848,0.280178); rgb(148pt)=(0.888315,0.382653,0.273942); rgb(149pt)=(0.892317,0.395453,0.267693); rgb(150pt)=(0.896132,0.408247,0.261429); rgb(151pt)=(0.899765,0.421034,0.255153); rgb(152pt)=(0.903217,0.433811,0.248865); rgb(153pt)=(0.906491,0.446577,0.242565); rgb(154pt)=(0.909591,0.459331,0.236255); rgb(155pt)=(0.91252,0.472069,0.229935); rgb(156pt)=(0.91528,0.484791,0.223606); rgb(157pt)=(0.917875,0.497493,0.217269); rgb(158pt)=(0.920308,0.510173,0.210923); rgb(159pt)=(0.922582,0.522828,0.204571); rgb(160pt)=(0.924702,0.535457,0.198211); rgb(161pt)=(0.926669,0.548055,0.191846); rgb(162pt)=(0.928489,0.560621,0.185476); rgb(163pt)=(0.930165,0.573152,0.179101); rgb(164pt)=(0.931701,0.585644,0.172723); rgb(165pt)=(0.9331,0.598094,0.166341); rgb(166pt)=(0.934367,0.6105,0.159958); rgb(167pt)=(0.935506,0.622859,0.153573); rgb(168pt)=(0.936521,0.635168,0.147187); rgb(169pt)=(0.937416,0.647424,0.140801); rgb(170pt)=(0.938196,0.659624,0.134415); rgb(171pt)=(0.938865,0.671766,0.128031); rgb(172pt)=(0.939428,0.683846,0.121649); rgb(173pt)=(0.939888,0.695862,0.115271); rgb(174pt)=(0.940251,0.707812,0.108895); rgb(175pt)=(0.940521,0.719693,0.102524); rgb(176pt)=(0.940702,0.731503,0.0961575); rgb(177pt)=(0.940799,0.743241,0.0897965); rgb(178pt)=(0.940816,0.754903,0.0834414); rgb(179pt)=(0.940758,0.766488,0.0770928); rgb(180pt)=(0.940628,0.777995,0.0707512); rgb(181pt)=(0.940431,0.789422,0.0644171); rgb(182pt)=(0.940172,0.800767,0.0580908); rgb(183pt)=(0.939853,0.81203,0.0517728); rgb(184pt)=(0.93948,0.823209,0.0454634); rgb(185pt)=(0.939055,0.834304,0.0391629); rgb(186pt)=(0.938583,0.845313,0.0328715); rgb(187pt)=(0.938067,0.856237,0.0265896); rgb(188pt)=(0.937511,0.867074,0.0203172); rgb(189pt)=(0.936918,0.877825,0.0140546); rgb(190pt)=(0.936291,0.888489,0.00780188); rgb(191pt)=(0.935633,0.899066,0.00155912); rgb(192pt)=(0.939757,0.904486,0.0115854); rgb(193pt)=(0.945337,0.908123,0.0270716); rgb(194pt)=(0.950752,0.911671,0.0426003); rgb(195pt)=(0.956001,0.91513,0.058169); rgb(196pt)=(0.961083,0.918506,0.0737758); rgb(197pt)=(0.965998,0.921799,0.0894185); rgb(198pt)=(0.970744,0.925013,0.105095); rgb(199pt)=(0.975322,0.92815,0.120804); rgb(200pt)=(0.979731,0.931211,0.136544); rgb(201pt)=(0.983971,0.9342,0.152312); rgb(202pt)=(0.988042,0.937117,0.168108); rgb(203pt)=(0.991945,0.939965,0.18393); rgb(204pt)=(0.995678,0.942744,0.199775); rgb(205pt)=(0.999244,0.945457,0.215644); rgb(206pt)=(0.993857,0.947511,0.269473); rgb(207pt)=(0.996135,0.950663,0.28468); rgb(208pt)=(0.998292,0.953724,0.299911); rgb(209pt)=(0.991889,0.954834,0.348602); rgb(210pt)=(0.993233,0.957982,0.363167); rgb(211pt)=(0.994497,0.961025,0.37775); rgb(212pt)=(0.995684,0.963963,0.392349); rgb(213pt)=(0.996793,0.966796,0.406962); rgb(214pt)=(0.997828,0.969525,0.421585); rgb(215pt)=(0.998789,0.972148,0.436217); rgb(216pt)=(0.999679,0.974668,0.450854); rgb(217pt)=(0.993936,0.974927,0.491285); rgb(218pt)=(0.994454,0.977293,0.505218); rgb(219pt)=(0.994938,0.979551,0.519149); rgb(220pt)=(0.995389,0.981703,0.533076); rgb(221pt)=(0.995808,0.983749,0.546997); rgb(222pt)=(0.996197,0.985689,0.560907); rgb(223pt)=(0.996557,0.987526,0.574806); rgb(224pt)=(0.996889,0.989259,0.588691); rgb(225pt)=(0.997196,0.990889,0.602559); rgb(226pt)=(0.997477,0.992419,0.616407); rgb(227pt)=(0.997736,0.993847,0.630234); rgb(228pt)=(0.997972,0.995177,0.644037); rgb(229pt)=(0.998187,0.996409,0.657813); rgb(230pt)=(0.998383,0.997545,0.67156); rgb(231pt)=(0.998561,0.998586,0.685277); rgb(232pt)=(0.994604,0.996743,0.712917); rgb(233pt)=(0.995201,0.99709,0.725975); rgb(234pt)=(0.995751,0.997412,0.738995); rgb(235pt)=(0.996257,0.99771,0.751973); rgb(236pt)=(0.993505,0.995985,0.774807); rgb(237pt)=(0.994069,0.996318,0.787149); rgb(238pt)=(0.994598,0.996633,0.799447); rgb(239pt)=(0.995095,0.99693,0.811697); rgb(240pt)=(0.995561,0.99721,0.823899); rgb(241pt)=(0.995998,0.997474,0.83605); rgb(242pt)=(0.996408,0.997723,0.848148); rgb(243pt)=(0.996792,0.997959,0.860191); rgb(244pt)=(0.997153,0.998181,0.872179); rgb(245pt)=(0.997492,0.998391,0.884109); rgb(246pt)=(0.99781,0.998589,0.895981); rgb(247pt)=(0.99811,0.998777,0.907792); rgb(248pt)=(0.998392,0.998956,0.919542); rgb(249pt)=(0.998658,0.999125,0.93123); rgb(250pt)=(0.99891,0.999287,0.942854); rgb(251pt)=(0.999149,0.999441,0.954414); rgb(252pt)=(0.999376,0.999588,0.96591); rgb(253pt)=(0.999593,0.99973,0.977339); rgb(254pt)=(0.999659,0.999774,0.989067); rgb(255pt)=(1,1,1)},
	colorbar,
	point meta min=-15,
	point meta max=15,
	    colorbar style={
		width=0.3cm,
		height=4cm,
		ytick={-15,-10,...,15},
		ylabel={\footnotesize{$10 \log_{10} \widehat{\mathit{CDR}}\,\mathrm{[dB]}$}},
	    }
	]
	\addplot [draw=none] coordinates {(0,0)};
	\end{axis}
\end{tikzpicture}
};

  \path
    ([shift={(-0,0)}]current bounding box.south west)
    ([shift={( 0, 2mm )}]current bounding box.north east);
\end{tikzpicture}
 		\vspace{-8mm}
    \caption{Coherent-to-diffuse power ratio estimates obtained from different estimators (columns) as a function of the complex spatial coherence estimate $\hat\Gamma_x$. The theoretical coherence of fully coherent ($\Gamma_s$) and fully diffuse ($\Gamma_n$) signals is marked by $\mathbf{\circ}$ and $\mathbf{\times}$, respectively, while the theoretical coherence of mixed signals lies on the connecting line. Estimators are computed using $\tilde\Gamma_s=\Gamma_s$, $\tilde\Gamma_n=\Gamma_n$. Parameters $d=8\,\mathrm{cm}$, $f=1\,\mathrm{kHz}$, different TDOAs (rows).}
    \label{fig:CDR}
\end{figure*}

\section{Coherent-to-Diffuse Power Ratio Estimation}

For most proposed postfilters, the gain function has been formulated directly as a function of auto- and cross-power spectral estimates \cite{simmer_post-filtering_2001,mccowan_microphone_2003}, which are typically obtained from the microphone signals by recursive averaging:
\begin{equation}
\label{eq:recursive}
\hat\Phi_{x_i x_j}(\frameix,f)=\lambda \hat\Phi_{x_i x_j}(\frameix-1,f) + (1-\lambda) X_i(\frameix,f) X_j^*(\frameix,f),
\end{equation}
where $\lambda$ is a constant between 0 and 1. 
We follow a different approach where we first derive an SNR estimate, which can then be used to apply any suppression technique such as the Wiener filter or spectral subtraction \cite{haensler_acoustic_2004}. Furthermore, we write the estimate not as a function of auto- and cross-power spectral estimates, but as a function of the estimated short-time spatial coherence, which allows additional insight into the behavior of the estimator. The short-time coherence is estimated by
\begin{equation}
\label{eq:coherence}
\hat\Gamma_x(\frameix,f)=\frac{\hat\Phi_{x_1 x_2}(\frameix,f)}{\sqrt{\hat\Phi_{x_1 x_1}(\frameix,f) \hat\Phi_{x_2 x_2}(\frameix,f)}}.
\end{equation}

Since the focus is on estimating the SNR for a mixture of a fully coherent signal with $|\Gamma_s(f)|=1$ and isotropic noise with $\Gamma_n \in \mathbb{R}$, where typically $\Gamma_n(f)=\Gamma_\text{diffuse}(f)$, we use the term CDR instead of SNR for the quantity to be estimated in the following.
For the application to dereverberation, the CDR is equivalent to the direct-to-reverberation power ratio (DRR), under the assumption that reverberant sound can be modeled as a mixture of a direct component and a perfectly diffuse reverberation component which are mutually uncorrelated, thus neglecting early reflections.

The aim is now to estimate the CDR from an estimate of the short-time spatial coherence $\hat\Gamma_x(\frameix,f)$, exploiting the known coherence functions of the signal and/or noise component, and the relationship of these coherence models and the mixed sound field coherence to the CDR given by (\ref{eq:Gamma_x_line}).
Solving (\ref{eq:Gamma_x_line}) for the CDR yields (for brevity, the time- and frequency-dependency is omitted in the following)
\begin{flalign}
\CDR &= \frac{\Gamma_n-\Gamma_x}{\Gamma_x-\Gamma_s},
\label{eq:CDR_ideal}
\end{flalign}
or, reformulated as the diffuseness $D$,
\begin{flalign}
D = \frac{1}{CDR+1} = \frac{\Gamma_x-\Gamma_s}{\Gamma_n-\Gamma_s}.
\end{flalign}
Although $\Gamma_x$ and $\Gamma_s$ may be complex, the CDR and diffuseness are real-valued quantities; however, when inserting a coherence estimate $\hat\Gamma_x$ for $\Gamma_x$ in (\ref{eq:CDR_ideal}), the resulting values are in general complex-valued, due to mismatch between the coherence models and the actual acoustic conditions, and the variance of the coherence estimate. Estimating the CDR by direct application of (\ref{eq:CDR_ideal}) is therefore not feasible, which is why a number of different estimator implementations, which yield a positive, real-valued CDR estimate for all possible values of $\hat\Gamma_x, |\hat\Gamma_x|\leq1$, have been proposed.

In the following, first, the interpretation of the estimator behavior in the complex plane is discussed. Then, existing and novel approaches to CDR estimation are analyzed. For an easier comparison, the estimators are reformulated as a function of only the coherence estimate $\hat\Gamma_x$ and the assumed coherence models $\tilde\Gamma_s$ and $\tilde\Gamma_n$, where $\tilde\Gamma_s$ is the direct signal coherence computed according to (\ref{eq:coh-direct}) from an a-priori known or estimated TDOA $\widehat{\Delta t}$, and $\tilde\Gamma_n$ is assumed to match the diffuse coherence model (\ref{eq:coh-diffuse}). We start with methods which make use of both $\tilde\Gamma_s$ and $\tilde\Gamma_n$, i.e., exploit information on the DOA and the noise coherence, continue with DOA-independent estimators which exploit only the knowledge of $\tilde\Gamma_n$, and finally propose a CDR estimator for the case of available signal coherence $\tilde\Gamma_s$, but unknown noise coherence. Table~\ref{table:overview} summarizes the presented estimators and their main properties. Finally, estimator bias and robustness are evaluated.

\subsection{Interpretation of Estimator Behavior in the Complex Plane}

Fig.~\ref{fig:CDR} shows the output of the estimators which are described in the following sections in the complex plane of possible coherence values $\hat\Gamma_x$. Results for a direct signal TDOA $\Delta t = 0$ (broadside) are shown in the first row, while in the second row, results are shown for $\Delta t = \frac{1}{5f}$. For all estimators, $\tilde\Gamma_s=\Gamma_s$, $\tilde\Gamma_n=\Gamma_n$ is assumed. The symbol $\mathbf{\circ}$ marks the coherence of a fully coherent signal with the respective TDOA according to (\ref{eq:coh-direct}), while the symbol $\mathbf{\times}$ marks the coherence of an ideal diffuse signal given by (\ref{eq:coh-diffuse}). The straight white line between these points marks the theoretical coherence values which would occur under ideal conditions for different CDR values, according to (\ref{eq:Gamma_x_line}). The \emph{bias} of a CDR estimator is henceforth defined as the deviation of the estimator from (\ref{eq:CDR_ideal}) for coherence values along this line; i.e., an \emph{unbiased} estimator should exactly match (\ref{eq:CDR_ideal}) for these values. This can be verified by inserting $\Gamma_x$ according to (\ref{eq:Gamma_x_line}) for $\hat\Gamma_x$ into the estimator equation, which yields $\widehat\CDR=\CDR$ for an unbiased estimator. Furthermore, since the coherence estimates $\hat\Gamma_x$, which are observed in practice, will not lie exactly on the line, a good estimator should also be \emph{robust} in the sense that some deviations of the coherence estimate from the assumed model, e.g., caused by an imperfect DOA estimate, do not lead to large deviations of the CDR estimate. In Fig.~\ref{fig:CDR}, robustness can be seen in the change of the CDR estimate for coherence values slightly deviating from the line; if these changes are abrupt, as in Fig.~\ref{fig:CDR}b for coherence values close to the unit circle, this indicates non-robust behavior. While we do not derive a measure for the overall robustness of an estimator, which would require establishing a statistical model for the errors, we evaluate the behavior of the different estimators with coherence model errors in Section~\ref{sec:bias-robustness-illustration}.
 
\begin{table}
	\centering
\caption{Overview of investigated CDR estimators, required prior information (noise and/or signal coherence) and unbiasedness.}
\label{table:overview}
    \begin{tabularx}{\columnwidth}{ l  l  l  l  X }
    \toprule
	Estimator & Definition & Required & Unbiased \\
	\midrule
    Jeub & $\frac{\tilde\Gamma_n-\Re \{\tilde\Gamma_s^* \hat\Gamma_x\}}{\Re \{\tilde\Gamma_s^* \hat\Gamma_x\}-1}$ & $\tilde\Gamma_n$, $\tilde\Gamma_s$ & no \\ \addlinespace[2pt]
    Thiergart 1 & $\Re \left\{ \frac{\tilde\Gamma_n-\hat\Gamma_x}{\hat\Gamma_x-\tilde\Gamma_s} \right\}$ & $\tilde\Gamma_n$, $\tilde\Gamma_s$ & yes \\ \addlinespace[2pt]
    Proposed 1 & $\frac{\Re \{\tilde\Gamma_s^* (\tilde\Gamma_n-\hat\Gamma_x)\}}{\Re \{\tilde\Gamma_s^* \hat\Gamma_x\}-1}$ & $\tilde\Gamma_n$, $\tilde\Gamma_s$ & yes \\ \addlinespace[2pt]
    Proposed 2 & $\frac{1-\tilde\Gamma_n \cos(\arg(\tilde\Gamma_s))}{|\tilde\Gamma_n - \tilde\Gamma_s|} \left| \frac{\tilde\Gamma_s^* (\tilde\Gamma_n-\hat\Gamma_x) }{\Re \{ \tilde\Gamma_s^* \hat\Gamma_{x}\}-1} \right|$ & $\tilde\Gamma_n$, $\tilde\Gamma_s$ & yes \\ \addlinespace[2pt]
	Thiergart 2 & $\Re \left\{ \frac{\tilde\Gamma_n-\hat\Gamma_x}{\hat\Gamma_x-e^{j \arg \hat\Gamma_x}} \right\}$ & $\tilde\Gamma_n$ & no \\ \addlinespace[2pt]
	Proposed 3 & (\ref{eq:prop3}) & $\tilde\Gamma_n$ & yes \\ \addlinespace[2pt]
	Proposed 4 & (\ref{eq:prop4}) & $\tilde\Gamma_s$ & yes \\ \addlinespace[2pt]
	\bottomrule
    \end{tabularx}
\end{table}

\subsection{CDR Estimation for Known DOA and Noise Coherence}

\begin{figure*}[b]
\hrulefill
\normalsize
\setcounter{MYtempeqncnt}{\value{equation}}
\setcounter{equation}{24}
\begin{equation}
\label{eq:prop3}
\widehat{\CDR}_\text{prop3}(\frameix,f) = \frac{\tilde\Gamma_n\, \Re\{\hat\Gamma_x\} -{|\hat\Gamma_x|}^2 - \sqrt{\tilde\Gamma_n^2\, {\Re\{\hat\Gamma_x\}}^2 - \tilde\Gamma_n^2\, {|\hat\Gamma_x|}^2 + \tilde\Gamma_n^2 - 2\, \tilde\Gamma_n\, \Re\{\hat\Gamma_x\} + {|\hat\Gamma_x|}^2}}{{|\hat\Gamma_x|}^2 - 1}
\end{equation}
\setcounter{equation}{\value{MYtempeqncnt}}
\end{figure*}

Using the same model as described in Section~\ref{sec:signal-model}, McCowan and Bourlard \cite{mccowan_microphone_2003} derived the Wiener postfilter for a coherent signal in diffuse noise. Jeub et al. \cite{jeub_model-based_2010} evaluated this postfilter for the suppression of reverberation, and formulated a CDR estimate based on the same model \cite{jeub_blind_2011}. Both McCowan and Jeub rely on the assumption that the direct signal is time-aligned in both microphones, which can be achieved by applying a delay corresponding to the TDOA estimate $\widehat{\Delta t}$ to one of the channels \cite{jeub_model-based_2010}. In the STFT domain, this delay is equivalent to a phase rotation of the cross-power spectrum (assuming that the delay is significantly shorter than the transform length), and can therefore be represented in the CDR estimator equation by multiplying the complex rotation factor $e^{-j 2 \pi f \widehat{\Delta t}}=\tilde\Gamma_s^*$ with the coherence estimate $\hat\Gamma_x$. This allows the formulation of the CDR estimator including time alignment as a function of only $\hat\Gamma_x$, $\tilde\Gamma_s$ and $\tilde\Gamma_n$:
\begin{flalign}
\label{eq:jeub}
\nonumber
\widehat{\CDR}_\text{Jeub}(\frameix,f)
&= \max\left(0,\frac{\tilde\Gamma_n-\Re \{e^{-j 2 \pi f \widehat{\Delta t}} \hat\Gamma_x\}}{\Re \{e^{-j 2 \pi f \widehat{\Delta t}} \hat\Gamma_x\}-1}\right)\\
&= \max\left(0,\frac{\tilde\Gamma_n-\Re \{\tilde\Gamma_s^* \hat\Gamma_x\}}{\Re \{\tilde\Gamma_s^* \hat\Gamma_x\}-1}\right).
\end{flalign}
The maximum operation is required to prevent negative results for the CDR estimate. This estimator is unbiased for $\tilde\Gamma_s=1$, i.e., $\widehat{\Delta t}=0$. However, for non-zero TDOAs, the phase rotation of the coherence estimate $\hat\Gamma_x$ does not only affect the direct signal component, but also the coherence of the diffuse signal component. Since this is not accounted for by this estimator, the estimate is biased for non-zero TDOAs. The estimator is illustrated in Fig.~\ref{fig:CDR}a.

Thiergart et al. \cite{thiergart_signal--reverberant_2012,thiergart_spatial_2012} proposed to estimate the CDR by directly inserting the target signal coherence estimate $\tilde\Gamma_s$ into (\ref{eq:CDR_ideal}), and taking the real part:
\begin{flalign}
\label{eq:thiergart1}
\widehat{\CDR}_\text{Thiergart1}(\frameix,f) &= \max\left(0,\Re \left\{ \frac{\tilde\Gamma_n-\hat\Gamma_x}{\hat\Gamma_x-\tilde\Gamma_s} \right\}\right).
\end{flalign}
While this estimator is unbiased, it was found to be very sensitive towards phase deviations of the coherence estimate from the ideal model \cite{thiergart_spatial_2012}. For a measured coherence with a magnitude close to one, even a small phase difference between $\hat\Gamma_x$ and $\Gamma_s$ can have a large effect on the CDR estimate.
This can be seen in Fig.~\ref{fig:CDR}b, where, unlike in Fig.~\ref{fig:CDR}a, the CDR for coherence values close to the unit circle sharply drops to zero, and is shown in more detail later.

Based on (\ref{eq:jeub}), an unbiased CDR estimator can be formulated \cite{schwarz_unbiased_2014}. The diffuse coherence model is first corrected to account for the phase rotation of the coherence estimate by multiplying the diffuse noise coherence $\tilde\Gamma_n$ with the phase term $e^{-j 2 \pi f \widehat{\Delta t}}$ as well, which removes the bias of the estimator, while preserving the robust properties of (\ref{eq:jeub}) against phase errors (see Fig.~\ref{fig:CDR}c):
\begin{flalign}
\label{eq:prop1}
\nonumber
\widehat{\CDR}_\text{prop1}(\frameix,f)
&= \max\left(0,\frac{\Re \{e^{-j 2 \pi f \widehat{\Delta t}} \tilde\Gamma_n-e^{-j 2 \pi f \widehat{\Delta t}} \hat\Gamma_x\}}{\Re \{e^{-j 2 \pi f \widehat{\Delta t}} \hat\Gamma_x\}-1}\right)\\
&= \max\left(0,\frac{\Re \{\tilde\Gamma_s^* (\tilde\Gamma_n-\hat\Gamma_x)\}}{\Re \{\tilde\Gamma_s^* \hat\Gamma_x\}-1}\right).
\end{flalign}
This estimator is identical to (\ref{eq:jeub}) for $\tilde\Gamma_s=1$, i.e., $\widehat{\Delta t}=0$.
Note that an equivalent CDR estimate can be derived from the maximum likelihood noise variance estimator which was proposed in \cite{ye_maximum_1995} and applied to noise reduction in \cite{kuklasinski_maximum_2014}.

For a second, heuristically motivated variant of an unbiased estimator, the real part in the numerator of (\ref{eq:prop1}) and the max operator are first replaced by the magnitude of the entire term. The resulting estimator was found to lead to an increased performance for the application to dereverberation \cite{schwarz_coherence-based_2014}:
\begin{flalign}
\label{eq:prop2_prime}
\widehat{\CDR}_\text{prop2}'(\frameix,f) &= \left| \frac{\tilde\Gamma_s^* (\tilde\Gamma_n-\hat\Gamma_x) }{\Re \{ \tilde\Gamma_s^* \hat\Gamma_{x}\}-1} \right|.
\end{flalign}
This estimator however has a small bias for non-zero TDOAs; a correction term for this bias can be computed by inserting (\ref{eq:Gamma_x_line}) into (\ref{eq:prop2_prime}) and solving for $\frac{\CDR}{\widehat{\CDR}_\text{prop2}'}$. The bias-compensated estimator is then given by
\begin{flalign}
\label{eq:prop2}
\widehat{\CDR}_\text{prop2}(\frameix,f) &= \frac{1-\tilde\Gamma_n \cos(\arg(\tilde\Gamma_s))}{|\tilde\Gamma_n - \tilde\Gamma_s|} \widehat{\CDR}_\text{prop2}'(\frameix,f),
\end{flalign}
and is illustrated in Fig.~\ref{fig:CDR}d. Compensation of this small bias however only has a negligible effect on practical performance. 

The derivation of these estimators shows that, when both knowledge of the signal and noise coherence are available, several different unbiased CDR estimators can be implemented. The reason for this is that the requirement of unbiasedness only defines the behavior of the estimator for coherence values matching the model given by (\ref{eq:Gamma_x_line}), i.e., the values on the line in Fig.~\ref{fig:CDR}, while allowing arbitrary behavior for other coherence values. While the second proposed unbiased variant has significant practical advantages, as shown in the qualitative analysis of the estimator behavior in Section~\ref{sec:bias-robustness-illustration} and the signal-based evaluation in Section~\ref{sec:evaluation}, it does not seem to be optimal in any sense. A possible direction for future work would therefore be to establish a statistical model for the deviations of $\hat\Gamma_x$ from the theoretical model given by (\ref{eq:Gamma_x_line}), and derive a correspondingly optimized unbiased estimator.

\subsection{CDR Estimation for Unknown DOA}

The previously shown methods rely on prior knowledge or an estimate of the target DOA. As an alternative, Thiergart et al. \cite{thiergart_signal--reverberant_2012,thiergart_spatial_2012} proposed to use the instantaneous phase of the estimated cross-power spectrum $\hat\Phi_{x_1 x_2}$ as a phase estimate for the direct signal model, i.e., $\tilde\Gamma_s=e^{j \arg \hat\Phi_{x_1 x_2}}$, thus removing the need for explicit DOA estimation to obtain $\tilde\Gamma_s$. Since, according to (\ref{eq:coherence}), $\arg \hat\Gamma_x=\arg \hat\Phi_{x_1 x_2}$, this estimator can be formulated as a function of only the coherence estimate $\hat\Gamma_x$ and the noise coherence $\tilde\Gamma_n$:
\begin{flalign}
\label{eq:thiergart2}
\widehat{\CDR}_\text{Thiergart2}(\frameix,f) &= \max \left(0, \Re \left\{ \frac{\tilde\Gamma_n-\hat\Gamma_x}{\hat\Gamma_x-e^{j \arg \hat\Gamma_x}} \right\}\right).
\end{flalign}
However, the instantaneous phase of the mixture is not an unbiased estimate of the phase of the direct signal component, since, for low CDR values, the coherence of the mixture is dominated by the coherence of the diffuse signal component \cite{thiergart_spatial_2012}, which is real-valued, i.e., has a phase of zero. For $\theta\neq0\,^\circ$, the estimator is therefore biased. The behavior of the estimator is illustrated in Fig.~\ref{fig:CDR}e.

As shown in \cite{schwarz_unbiased_2014}, it is possible to derive an unbiased CDR estimator which does not require an estimate of the source DOA, since the knowledge that $|\Gamma_s|=1$, i.e., that the direct signal is fully coherent, is sufficient to solve (\ref{eq:CDR_ideal}). This can be explained using a geometric interpretation: according to (\ref{eq:Gamma_x_line}), $\Gamma_x$, $\Gamma_s$ and $\Gamma_n$ all lie on a straight line in the complex plane, and it is furthermore known that $\Gamma_s$ lies on the unit circle and $\Gamma_n$ on the real axis. $\Gamma_s$ can therefore be obtained by the intersection of the line through $\Gamma_n$ and $\Gamma_x$ with the unit circle, and inserted into (\ref{eq:CDR_ideal}). An alternative way of obtaining this solution is by solving (\ref{eq:Gamma_x_line}) for $\Gamma_s$ and setting the magnitude to 1:
\begin{flalign}
|\Gamma_s| = \left|\Gamma_x-(\Gamma_n-\Gamma_x) \CDR^{-1}\right| \overset{!}{=} 1, %
\end{flalign}
which leads to a quadratic equation for the CDR:
\begin{align}
\nonumber
(|\Gamma_x|^2-1) \CDR^2 - 2 \Re\{ \Gamma_x(\Gamma_n-\Gamma_x)^* \} \CDR \\ +|\Gamma_n-\Gamma_x|^2 = 0.
\end{align}%
\addtocounter{equation}{1}%
Taking the positive of both possible solutions yields the unbiased DOA-independent CDR estimator which is given by (\ref{eq:prop3}) and illustrated in Fig.~\ref{fig:CDR}f. In contrast to the DOA-dependent estimators, where an infinite number of unbiased estimators exists, the DOA-independent estimator is uniquely determined by the requirement of unbiasedness.

\subsection{CDR Estimation for Unknown Noise Coherence}

From the geometric interpretation of the coherence of mixed sound fields it can be analogously concluded that knowledge of $\Gamma_n$ is not required when $\Gamma_s$ is known, since the noise coherence is assumed to be real and therefore determined by the intersection of the real axis and the line through $\Gamma_s$ and $\Gamma_x$. Using $\Im\{\Gamma_n\}=0$, $\Gamma_n$ can therefore be eliminated from (\ref{eq:CDR_ideal}), resulting in
\begin{equation}
\CDR = \frac{\Im \{\Gamma_x\}}{\Im \{\Gamma_s\}-\Im \{\Gamma_x\}}.
\end{equation}
When using this formulation with the estimates $\hat\Gamma_x$ and $\tilde\Gamma_s$ as an estimator for the CDR, practical problems occur in cases where, due to model mismatch and coherence estimation errors, the imaginary part of the coherence estimate $\Im\{\hat\Gamma_x\}$ has either values with a larger magnitude than $\Im\{\tilde\Gamma_s\}$, or a different sign, in which case this equation would not yield a meaningful result. For this reason, the CDR estimate is continuously extended into these two problematic regions by returning an infinite CDR in the former case, and a CDR of zero in the latter case. The final proposed estimator is then given by
\begin{equation}
\label{eq:prop4}
\widehat{\CDR}_\text{prop4}(\frameix,f) =
\begin{cases}
 \infty, & \text{for } \frac{\Im \{\hat\Gamma_x\}}{\Im \{\tilde\Gamma_s\}} \geq 1 \\
 \frac{\Im \{\hat\Gamma_x\}}{\Im \{\tilde\Gamma_s\}-\Im \{\hat\Gamma_x\}}, & \text{for } 0 < \frac{\Im \{\hat\Gamma_x\}}{\Im \{\tilde\Gamma_s\}} < 1 \\
 0, & \text{for } \frac{\Im \{\hat\Gamma_x\}}{\Im \{\tilde\Gamma_s\}} \leq 0.
\end{cases}
\end{equation}
An inherent constraint that limits practical applicability of this estimator is that $\arg \Gamma_s \neq 0$, since otherwise the imaginary parts disappear; i.e., the estimator is not usable for $\Delta t = 0$, and increasingly sensitive towards estimation errors for small TDOAs. The estimator is visualized in Fig.~\ref{fig:CDR}g. Note that in \cite{ito_designing_2010} a noise power spectrum estimate was derived in a similar way from the imaginary part of a cross-power spectrum. 

\subsection{Evaluation of Estimator Bias and Robustness}
\label{sec:bias-robustness-illustration}
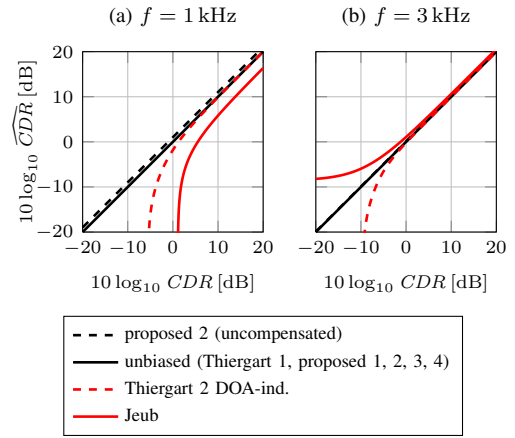
\begin{figure}[t]
    \centering
	\setlength\figureheight{2.4cm}
	\setlength\figurewidth{2.4cm}
	\pgfplotsset{
	title style={font=\footnotesize},
	tick label style={font=\scriptsize},
	label style={font=\scriptsize},
	legend style={font=\scriptsize,align=left},
	}
\definecolor{mycolor1}{rgb}{0.00000,0.50000,1.00000}%
\begin{tikzpicture}
\begin{groupplot}[%
group style={
	columns=2,
	rows=1,
	xlabels at=edge bottom,
	ylabels at=edge left,
	horizontal sep=7mm,
	group name=plots
},
height=\figureheight,
width=\figurewidth,
scale only axis,
legend columns=1,
legend cell align={left},
legend style={align=left},
y label style={at={(-0.2,0.5)}}
]
\nextgroupplot[%
unbounded coords=jump,
scale only axis,
xmin=-20,
xmax=20,
xlabel={$10 \log_{10} \mathit{CDR}\,\mathrm{[dB]}$},
xmajorgrids,
ymin=-20,
ymax=20,
ylabel={$10 \log_{10} \widehat{\mathit{CDR}}\,\mathrm{[dB]}$},
ymajorgrids,
title={(a) $f=1\,\mathrm{kHz}$}
]
\addplot [line width=1.0pt,color=black,dashed] table[row sep=crcr]{figures/estimator-bias-comparison-f1000-1.tsv};
\addplot [line width=1.0pt,color=black,solid] table[row sep=crcr]{figures/estimator-bias-comparison-f1000-2.tsv};
\addplot [line width=1.0pt,color=red,dashed] table[row sep=crcr]{figures/estimator-bias-comparison-f1000-3.tsv};
\addplot [line width=1.0pt,color=red,solid]  table[row sep=crcr]{figures/estimator-bias-comparison-f1000-4.tsv};
\nextgroupplot[%
unbounded coords=jump,
scale only axis,
xmin=-20,
xmax=20,
xlabel={$10 \log_{10} \mathit{CDR}\,\mathrm{[dB]}$},
xmajorgrids,
ymin=-20,
ymax=20,
ymajorgrids,
yticklabels={\empty},
legend to name=biaslegend,
title={(b) $f=3\,\mathrm{kHz}$}
]
\addplot [line width=1.0pt,color=black,dashed]
  table[row sep=crcr]{figures/estimator-bias-comparison-f3000-1.tsv};
\addlegendentry{proposed 2 (uncompensated)}
\addplot [line width=1.0pt,color=black,solid] table[row sep=crcr]{figures/estimator-bias-comparison-f3000-2.tsv};
\addlegendentry{unbiased (Thiergart 1, proposed 1, 2, 3, 4)}
\addplot [line width=1.0pt,color=red,dashed] table[row sep=crcr]{figures/estimator-bias-comparison-f3000-3.tsv};
\addlegendentry{Thiergart 2 DOA-ind.}
\addplot [line width=1.0pt,color=red,solid] table[row sep=crcr]{figures/estimator-bias-comparison-f3000-4.tsv};
\addlegendentry{Jeub}
\end{groupplot}
\node at (plots c1r1.south east) [inner sep=0pt,anchor=north, yshift=-7ex] {\pgfplotslegendfromname{biaslegend}};
\end{tikzpicture}%
     \caption{Comparison of true CDR and estimated CDR. Parameters $d=8\,\mathrm{cm}$, $\widehat{\Delta t} = \Delta t = \frac{1}{5f}$, $f=1\,\mathrm{kHz}$ (left), $3\,\mathrm{kHz}$ (right).}
    \label{fig:CDR_bias}
\end{figure}

\begin{figure}[t]
    \centering
	\setlength\figureheight{2.4cm}
	\setlength\figurewidth{2.4cm}
	\pgfplotsset{
	title style={font=\footnotesize},
	tick label style={font=\scriptsize},
	label style={font=\scriptsize},
	legend style={font=\scriptsize,align=left},
	}
\definecolor{mycolor1}{rgb}{0.00000,0.50000,1.00000}%
\begin{tikzpicture}
\begin{groupplot}[%
group style={
	columns=2,
	rows=2,
	xlabels at=edge bottom,
	ylabels at=edge left,
	horizontal sep=7mm,
	vertical sep=18mm,
	group name=plotsrobust
},
height=\figureheight,
width=\figurewidth,
scale only axis,
legend columns=2,
legend cell align={left},
legend style={align=left},
y label style={at={(-0.2,0.5)}}
]
\nextgroupplot[%
unbounded coords=jump,
scale only axis,
xmin=-0.3,
xmax=0.3,
xlabel={$\tilde \Gamma_n-\Gamma_n$},
xmajorgrids,
ymin=-10,
ymax=10,
ylabel={$10 \log_{10} \widehat{\mathit{CDR}} - 10 \log_{10} \mathit{CDR}$},
ymajorgrids,
title={(a) $\mathit{CDR}=-10\,\mathrm{dB}$}
]
\addplot [color=black,solid,line width=1.0pt]
  table[row sep=crcr]{figures/estimator-robustness-comparison-1.tsv};
\addplot [color=mycolor1,dash pattern=on 1pt off 3pt on 3pt off 3pt,line width=1.0pt]
  table[row sep=crcr]{figures/estimator-robustness-comparison-2.tsv};
\addplot [color=red,dashed,line width=1.0pt]
  table[row sep=crcr]{figures/estimator-robustness-comparison-3.tsv};
\addplot [color=black!20!green,dotted,line width=1.0pt]
  table[row sep=crcr]{figures/estimator-robustness-comparison-4.tsv};
\addplot [color=black,only marks,mark=diamond,mark options={solid}]
  table[row sep=crcr]{figures/estimator-robustness-comparison-5.tsv};
\nextgroupplot[%
unbounded coords=jump,
scale only axis,
xmin=-1.5707963267949,
xmax=1.5707963267949,
xlabel={$\arg \tilde \Gamma_s-\arg \Gamma_s$ $[$rad$]$},
xmajorgrids,
ymin=-10,
ymax=10,
ymajorgrids,
title={(b) $\mathit{CDR}=-10\,\mathrm{dB}$}
]
\addplot [color=black,solid,line width=1.0pt]
  table[row sep=crcr]{figures/estimator-robustness-comparison-6.tsv};
\addplot [color=mycolor1,dash pattern=on 1pt off 3pt on 3pt off 3pt,line width=1.0pt]
  table[row sep=crcr]{figures/estimator-robustness-comparison-7.tsv};
\addplot [color=red,dashed,line width=1.0pt]
  table[row sep=crcr]{figures/estimator-robustness-comparison-8.tsv};
\addplot [color=black!20!green,dotted,line width=1.0pt]
  table[row sep=crcr]{figures/estimator-robustness-comparison-9.tsv};
\addplot [color=black,only marks,mark=diamond,mark options={solid}]
  table[row sep=crcr]{figures/estimator-robustness-comparison-10.tsv};
\nextgroupplot[%
unbounded coords=jump,
scale only axis,
xmin=-0.3,
xmax=0.3,
xlabel={$\tilde \Gamma_n-\Gamma_n$},
xmajorgrids,
ymin=-10,
ymax=10,
ylabel={$10 \log_{10} \widehat{\mathit{CDR}} - 10 \log_{10} \mathit{CDR}$},
ymajorgrids,
title={(c) $\mathit{CDR}=10\,\mathrm{dB}$}
]
\addplot [color=black,solid,line width=1.0pt]
  table[row sep=crcr]{figures/estimator-robustness-comparison-16.tsv};
\addplot [color=mycolor1,dash pattern=on 1pt off 3pt on 3pt off 3pt,line width=1.0pt]
  table[row sep=crcr]{figures/estimator-robustness-comparison-17.tsv};
\addplot [color=red,dashed,line width=1.0pt]
  table[row sep=crcr]{figures/estimator-robustness-comparison-18.tsv};
\addplot [color=black!20!green,dotted,line width=1.0pt]
  table[row sep=crcr]{figures/estimator-robustness-comparison-19.tsv};
\addplot [color=black,only marks,mark=diamond,mark options={solid}]
  table[row sep=crcr]{figures/estimator-robustness-comparison-20.tsv};
\nextgroupplot[%
unbounded coords=jump,
scale only axis,
xmin=-1.5707963267949,
xmax=1.5707963267949,
xlabel={$\arg \tilde \Gamma_s-\arg \Gamma_s$ $[$rad$]$},
xmajorgrids,
ymin=-10,
ymax=10,
ymajorgrids,
legend to name=legend1,
title={(d) $\mathit{CDR}=10\,\mathrm{dB}$}
]
\addplot [color=black,solid,line width=1.0pt]
  table[row sep=crcr]{figures/estimator-robustness-comparison-11.tsv};
\addlegendentry{proposed 2}
\addplot [color=mycolor1,dash pattern=on 1pt off 3pt on 3pt off 3pt,line width=1.0pt]
  table[row sep=crcr]{figures/estimator-robustness-comparison-12.tsv};
\addlegendentry{proposed 1};
\addplot [color=red,dashed,line width=1.0pt]
  table[row sep=crcr]{figures/estimator-robustness-comparison-13.tsv};
\addlegendentry{Thiergart 1};
\addplot [color=black!20!green,dotted,line width=1.0pt]
  table[row sep=crcr]{figures/estimator-robustness-comparison-14.tsv};
\addlegendentry{proposed 3};
\addplot [color=black,only marks,mark=diamond,mark options={solid}]
  table[row sep=crcr]{figures/estimator-robustness-comparison-15.tsv};
\end{groupplot}
\node at (plotsrobust c1r2.south east) [inner sep=0pt,anchor=north, yshift=-7ex] {\pgfplotslegendfromname{legend1}};
\end{tikzpicture}
     \caption{CDR estimation error for noise and direct signal coherence model errors. Parameters $d=8\,\mathrm{cm}$, $\widehat{\Delta t} = \frac{1}{5f}$, $f=1\,\mathrm{kHz}$.}
    \label{fig:CDR_robustness}
\end{figure}
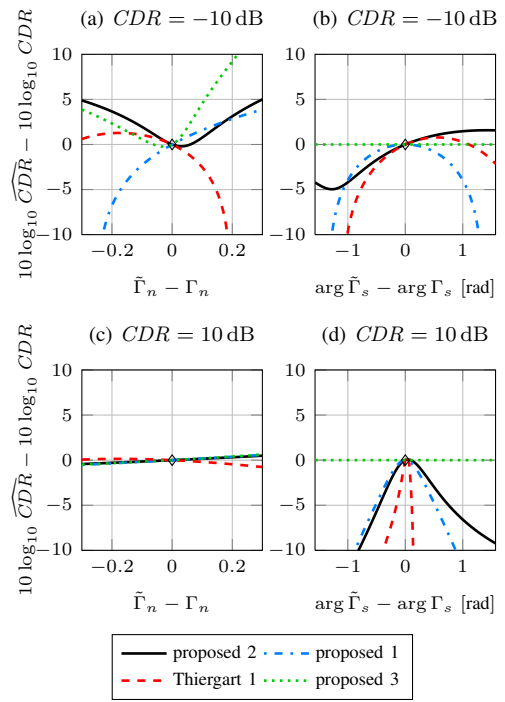

To illustrate the bias of the estimators $\CDRjeub$, the uncompensated estimator $\CDRproptwo'$ and $\CDRthiergarttwo$, Fig.~\ref{fig:CDR_bias} compares the true CDR value and the different estimates for mixtures of coherent and ideally diffuse signals for a TDOA $\Delta t=\frac{1}{5f}$ (corresponding to the values along the white line in Fig.~\ref{fig:CDR}, second row). The proposed estimators are all unbiased, as is the DOA-dependent estimator proposed by Thiergart et al. (\ref{eq:thiergart1}). The estimator by Jeub et al. (\ref{eq:jeub}) and the DOA-independent estimator by Thiergart et al. (\ref{eq:thiergart2}) both have a significant bias, with the former under- or overestimating the CDR depending on the values of $\Delta t$ and $f$, and the latter always underestimating the CDR. Also shown is the uncompensated version of the proposed estimator 2 (\ref{eq:prop2_prime}), which has a small, TDOA- and frequency-dependent bias (for $f=3\,\mathrm{kHz}$, the difference to the unbiased case is too small to be noticeable in the plot).

Fig.~\ref{fig:CDR_robustness} shows the CDR estimation error for cases where the actual coherence of the noise $\Gamma_n$ or the direct signal component $\Gamma_s$ deviates from the assumed coherence models $\tilde\Gamma_n$ and $\tilde\Gamma_s$, respectively. Fig.~\ref{fig:CDR_robustness}a and b show the error for a low CDR of $-10\,\dB$, while c and d show results for a high CDR of $10\,\dB$. The DOA-independent estimator $\CDRpropthree$ is naturally unaffected by the phase error of the direct signal coherence model, as seen in Fig.~\ref{fig:CDR_robustness}b and d; however, for errors of the noise coherence, the CDR is quickly overestimated by the DOA-independent estimator (see Fig.~\ref{fig:CDR_robustness}a).  The estimator $\CDRthiergartone$ has the problem of reacting strongly to small phase deviations when the CDR is high (see Fig.~\ref{fig:CDR_robustness}d). Comparing the different unbiased DOA-dependent variants $\CDRpropone$ and $\CDRproptwo$, it can be stated that $\CDRproptwo$ seems slightly more tolerant towards model errors, which could explain the better performance of this estimator for signal enhancement.

\section{Application to Speech Enhancement}
\label{sec:ReverberationSuppression}

\begin{figure}[tb]
    \centering
    \includegraphics[width=\columnwidth]{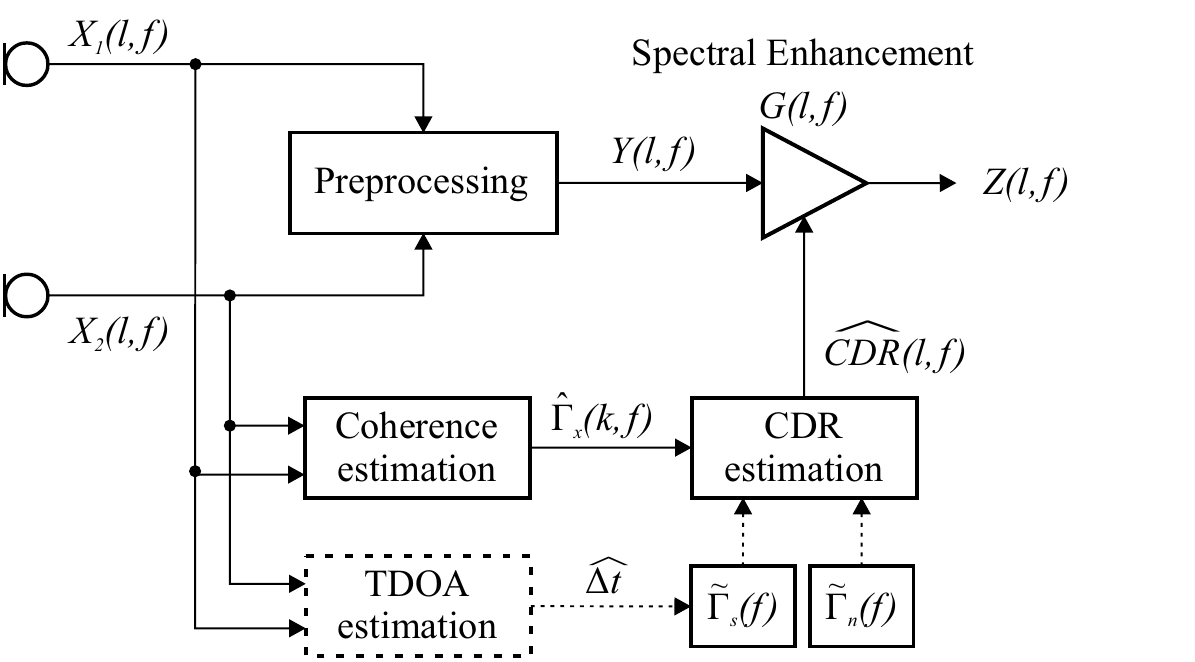}
    \caption{Coherence-based noise and reverberation suppression system consisting of a preprocessor and a CDR-based postfilter.}
    \label{fig:system}
\end{figure}

Fig.~\ref{fig:system} shows the structure of the proposed reverberation or diffuse noise suppression system based on short-time CDR estimates.
First, the microphone signals are combined by averaging the squared magnitudes and using the phase from one of the microphone signals:
\begin{align}
\label{eq:psdavg}
Y(\frameix,f)=&\frac{1}{2}\sqrt{|X_1(\frameix,f)|^2 + |X_2(\frameix,f)|^2} \cdot e^{j\arg X_{1}(\frameix,f)}.
\end{align}
Spatial magnitude averaging in the STFT domain is typically used to reduce the variance of spectral estimates for the computation
of microphone array postfilters \cite{mccowan_microphone_2003}, but has also been used as a preprocessor for signal enhancement \cite{habets_single-_2007}. It is used here with the purpose of reducing the variations in the transfer function which are caused by constructive and destructive interference of early reflection components with the direct path.
For the computation of the coherence-based postfilter gain $G(\frameix,f)$, short-time estimates $\hat{\Gamma}_x(\frameix,f)$ of the spatial coherence are first obtained according to (\ref{eq:coherence}) from spectra which have been estimated by recursive averaging. From the coherence, the CDR is estimated based on models for the direct signal and/or reverberation coherence, where the direct signal coherence is derived from a known or estimated TDOA, and the reverberation coherence is assumed to be known. A postfilter gain is then computed using spectral magnitude subtraction \cite{haensler_acoustic_2004}:
\begin{flalign}
G(\frameix,f) &= \max\left\{ G_\mathrm{min}, 1-\sqrt{\frac{\mu}{\widehat \CDR(\frameix,f)+1}} \right\},
\label{eq:Magnitude_Subtraction}
\end{flalign}
with the oversubtraction factor $\mu$ and the gain floor $G_\mathrm{min}$. The output signal is computed by applying the postfilter gain to the preprocessed signal $Y(\frameix,f)$, i.e., $Z(\frameix,f)=G(\frameix,f) Y(\frameix,f)$, and transformed back into the time domain.
Since the preprocessor does not have any spatial filtering effect, the postfilter gain can be directly applied to the preprocessor output, and does not require a correction to account for spatial filtering, as it would be the case for a beamformer as preprocessor \cite{simmer_post-filtering_2001}.

Note that, when employing a DOA-independent CDR estimator, the proposed signal enhancement system is completely independent of the DOA of the target signal.

\section{Evaluation}
\label{sec:evaluation}
In the following, the spatial properties of reverberation are first evaluated using simulated and measured RIRs, in order to verify the assumptions made in Sect.~\ref{sec:reverberation-coherence}. Then, the estimation accuracy of the CDR estimators and the effect of the proposed CDR-based dereverberation system are evaluated.

A MATLAB implementation of the proposed CDR estimators and signal enhancement scheme is provided online\footnote{\url{http://www.lms.lnt.de/files/publications/cdr-dereverb.zip}}.

\subsection{Setup and Parameters}

For the main evaluation, sets of measured RIRs from three rooms are used:
\begin{itemize}
\item Room A: $6\,\mathrm{m}\times6\,\mathrm{m}\times3\,\mathrm{m}$, partially closed curtains on walls, $T_{60}\approx0.4\,\mathrm{s}$
\item Room B: $7\,\mathrm{m}\times11\,\mathrm{m}\times3\,\mathrm{m}$ (lecture hall),
$T_{60}\approx1\,\mathrm{s}$
\item Room C: $54\,\mathrm{m}\times7\,\mathrm{m}\times3\,\mathrm{m}$ (large foyer).
$T_{60}\approx3.5\,\mathrm{s}$
\end{itemize}
The reverberation time $T_{60}$ was measured from the energy decay curve of the RIR.
In each room, RIRs were measured for 40-70 different source positions in $l=1$, $2$ and $4\,\mathrm{m}$ distance from the microphones, in the angular range $\theta=-90\dots90\,^\circ$. Microphones are spaced $d=$8\,cm apart.

Additionally, the RIRs that were used in the REVERB challenge \cite{kinoshita_reverb_2013} for the generation of multi-condition training data are evaluated. These RIRs were measured using an 8-channel circular microphone array with a diameter of 20\,cm (corresponding to $d=8\,\mathrm{cm}$ spacing between neighboring microphones) in 6 different rooms (SR1/2, MR1/2, LR1/2), for two source-microphone distances ($\approx$0.5\,m and $\approx$2\,m), and two different angles of the source w.r.t. the microphone array.
The rooms have the following properties (note that SR2 and LR2 are the same rooms as A and B, respectively):
\begin{itemize}
\item SR1 (``Small Room 1''): variable reverberation room, $4.5\,\mathrm{m}\times3.5\,\mathrm{m}\times3\,\mathrm{m}$, $T_{60}\approx0.2\,\mathrm{s}$
\item SR2 (``Small Room 2''): room A, but curtains fully closed, $T_{60}\approx0.2\,\mathrm{s}$
\item MR1 (``Medium Room 1''): same as SR1, $T_{60}\approx0.5\,\mathrm{s}$
\item MR2 (``Medium Room 2''): meeting room, $5\,\mathrm{m}\times3.5\,\mathrm{m}\times3\,\mathrm{m}$, $T_{60}\approx0.6\,\mathrm{s}$
\item LR1 (``Large Room 1''): same as SR1, $T_{60}\approx0.8\,\mathrm{s}$
\item LR2 (``Large Room 2''): room B
\end{itemize}

In the following, all processing takes place at a sampling rate of 16\,kHz. For the transformation into the time-frequency domain and short-time spectral estimation, a DFT-based uniform filterbank with window length 1024, FFT size 512, and downsampling factor 128 is employed \cite{harteneck_design_1999}. The short-time coherence estimates are obtained by recursive averaging of the auto- and cross-power spectra according to (\ref{eq:recursive}), with the forgetting factor $\lambda=0.68$.

\subsection{Spatial Properties of Reverberation in Simulated and Measured Rooms}

\label{sec:evaluation-reverberation-coherence}
For the evaluation of the spatial characteristics of reverberation, we use simulated and measured RIRs. The reverberation tail of the RIRs is extracted by removing the initial part containing the direct path and early reflections (see Appendix), using a typical value of $T_\mathrm{e}=50\,\mathrm{ms}$ for the cutoff time between early reflections and reverberation \cite{kuttruff_room_2000}. The late RIRs are convolved with a speech signal, transformed into the STFT domain, and the spatial coherence is estimated from auto- and cross-power spectra estimated by averaging over an interval of 10\,s.

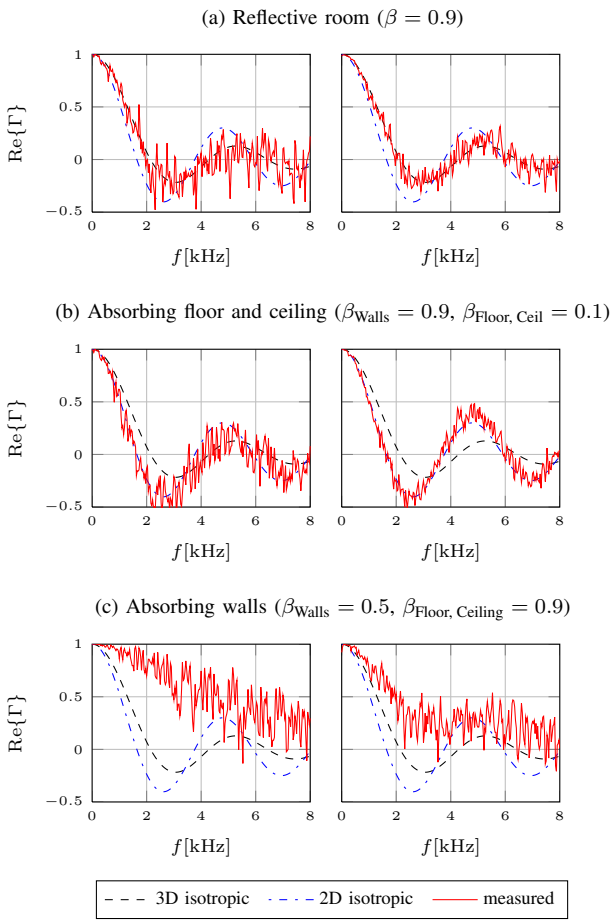
\begin{figure}[t]
    \centering
	\setlength\figureheight{2.1cm}
	\setlength\figurewidth{2.9cm}
	\pgfplotsset{
	tick label style={font=\tiny},
	label style={font=\scriptsize},
	legend style={font=\scriptsize,row sep=-1mm},
	}
	\subfloat[Reflective room ($\beta=0.9$)]{%
\begin{tikzpicture}

\begin{axis}[%
width=\figurewidth,
height=\figureheight,
scale only axis,
xmin=0,
xmax=8,
xlabel={$f \mathrm{[kHz]}$},
xmajorgrids,
ymin=-0.5,
ymax=1,
ylabel={$\Re\{\Gamma\}$},
ymajorgrids,
]
\addplot [color=black,dashed]
  table[row sep=crcr]{figures/coherence-simulated-rir-1-1.tsv};

\addplot [color=blue,dash pattern=on 1pt off 3pt on 3pt off 3pt]
  table[row sep=crcr]{figures/coherence-simulated-rir-1-2.tsv};

\addplot [color=red,solid]
  table[row sep=crcr]{figures/coherence-simulated-rir-1-3.tsv};

\end{axis}
\end{tikzpicture}%
\begin{tikzpicture}

\begin{axis}[%
width=\figurewidth,
height=\figureheight,
scale only axis,
xmin=0,
xmax=8,
xlabel={$f \mathrm{[kHz]}$},
xmajorgrids,
ymin=-0.5,
ymax=1,
ytick={-0.5,0,0.5,1},
yticklabels={\empty},
ymajorgrids,
]
\addplot [color=black,dashed]
  table[row sep=crcr]{figures/coherence-simulated-rir-2-1.tsv};

\addplot [color=blue,dash pattern=on 1pt off 3pt on 3pt off 3pt]
  table[row sep=crcr]{figures/coherence-simulated-rir-2-2.tsv};

\addplot [color=red,solid]
  table[row sep=crcr]{figures/coherence-simulated-rir-2-3.tsv};

\end{axis}
\end{tikzpicture}%
\hspace{11.2mm}
	}\hfil
	\subfloat[Absorbing floor and ceiling ($\beta_\text{Walls} = 0.9$, $\beta_\text{Floor, Ceil} = 0.1$)]{%
\begin{tikzpicture}

\begin{axis}[%
width=\figurewidth,
height=\figureheight,
scale only axis,
xmin=0,
xmax=8,
xlabel={$f \mathrm{[kHz]}$},
xmajorgrids,
ymin=-0.5,
ymax=1,
ylabel={$\Re\{\Gamma\}$},
ymajorgrids,
]
\addplot [color=black,dashed]
  table[row sep=crcr]{figures/coherence-simulated-rir-3-1.tsv};

\addplot [color=blue,dash pattern=on 1pt off 3pt on 3pt off 3pt]
  table[row sep=crcr]{figures/coherence-simulated-rir-3-2.tsv};

\addplot [color=red,solid]
  table[row sep=crcr]{figures/coherence-simulated-rir-3-3.tsv};

\end{axis}
\end{tikzpicture}%
\begin{tikzpicture}

\begin{axis}[%
width=\figurewidth,
height=\figureheight,
scale only axis,
xmin=0,
xmax=8,
xlabel={$f \mathrm{[kHz]}$},
xmajorgrids,
ymin=-0.5,
ymax=1,
ytick={-0.5,0,0.5,1},
yticklabels={\empty},
ymajorgrids,
]
\addplot [color=black,dashed]
  table[row sep=crcr]{figures/coherence-simulated-rir-4-1.tsv};

\addplot [color=blue,dash pattern=on 1pt off 3pt on 3pt off 3pt]
  table[row sep=crcr]{figures/coherence-simulated-rir-4-2.tsv};

\addplot [color=red,solid]
  table[row sep=crcr]{figures/coherence-simulated-rir-4-3.tsv};

\end{axis}
\end{tikzpicture}%
\hspace{11.2mm}
	}\hfil
	\subfloat[Absorbing walls ($\beta_\text{Walls} = 0.5$, $\beta_\text{Floor, Ceiling} = 0.9$)]{%
\begin{tikzpicture}

\begin{axis}[%
width=\figurewidth,
height=\figureheight,
scale only axis,
xmin=0,
xmax=8,
xlabel={$f \mathrm{[kHz]}$},
xmajorgrids,
ymin=-0.5,
ymax=1,
ylabel={$\Re\{\Gamma\}$},
ymajorgrids,
]
\addplot [color=black,dashed]
  table[row sep=crcr]{figures/coherence-simulated-rir-5-1.tsv};

\addplot [color=blue,dash pattern=on 1pt off 3pt on 3pt off 3pt]
  table[row sep=crcr]{figures/coherence-simulated-rir-5-2.tsv};

\addplot [color=red,solid]
  table[row sep=crcr]{figures/coherence-simulated-rir-5-3.tsv};

\end{axis}
\end{tikzpicture}%
\begin{tikzpicture}

\begin{axis}[%
width=\figurewidth,
height=\figureheight,
scale only axis,
xmin=0,
xmax=8,
xlabel={$f \mathrm{[kHz]}$},
xmajorgrids,
ymin=-0.5,
ymax=1,
ytick={-0.5,0,0.5,1},
yticklabels={\empty},
ymajorgrids,
]
\addplot [color=black,dashed]
  table[row sep=crcr]{figures/coherence-simulated-rir-6-1.tsv};

\addplot [color=blue,dash pattern=on 1pt off 3pt on 3pt off 3pt]
  table[row sep=crcr]{figures/coherence-simulated-rir-6-2.tsv};

\addplot [color=red,solid]
  table[row sep=crcr]{figures/coherence-simulated-rir-6-3.tsv};

\end{axis}
\end{tikzpicture}%
\hspace{11.2mm}
	}\vspace{2mm}\hfil\\
\begin{tikzpicture}
    \begin{customlegend}[legend entries={3D isotropic,2D isotropic,measured},legend columns=3,legend style={align=left,/tikz/every even column/.append style={column sep=5pt}},legend cell align={left}]
    \addlegendimage{color=black,dashed}
    \addlegendimage{color=blue,dash pattern=on 1pt off 3pt on 3pt off 3pt}
    \addlegendimage{color=red,solid}
    \end{customlegend}
\end{tikzpicture}
     \caption{Spatial coherence estimated from the reverberation tail of simulated RIRs, averaged over 7 microphone pairs with spacing $d=8\,\mathrm{cm}$, for different reflection coefficients $\beta$, compared to coherence of diffuse and 2D isotropic sound fields. Left: small room ($4\times3\times2.5\,$m), right: large room ($15\times18\times10\,$m).}
    \label{fig:coherence-simulated-rooms}
\end{figure}

\begin{figure}[t]
    \centering
	\setlength\figureheight{2.1cm}
	\setlength\figurewidth{2.9cm}
\pgfplotsset{
tick label style={font=\tiny},
label style={font=\scriptsize},
legend style={font=\scriptsize,row sep=-1mm},
}
	\subfloat[Small room 1]{\hspace{-11.2mm}%
\begin{tikzpicture}

\begin{axis}[%
width=\figurewidth,
height=\figureheight,
scale only axis,
xmin=0,
xmax=8,
xlabel={$f \mathrm{[kHz]}$},
xmajorgrids,
ymin=-0.5,
ymax=1,
ylabel={$\Re\{\Gamma\}$},
ymajorgrids,
yticklabel style={overlay} %
]
\addplot [color=black,dashed]
  table[row sep=crcr]{figures/coherence-measured-rir-1-1.tsv};

\addplot [color=blue,dash pattern=on 1pt off 3pt on 3pt off 3pt]
  table[row sep=crcr]{figures/coherence-measured-rir-1-2.tsv};

\addplot [color=red,solid]
  table[row sep=crcr]{figures/coherence-measured-rir-1-3.tsv};

\end{axis}

  \path
    ([shift={(-0,0)}]current bounding box.south west)
    ([shift={( 0, 1mm )}]current bounding box.north east);

\end{tikzpicture}%
}
	\subfloat[Small room 2]{%
\begin{tikzpicture}

\begin{axis}[%
width=\figurewidth,
height=\figureheight,
scale only axis,
xmin=0,
xmax=8,
xlabel={$f \mathrm{[kHz]}$},
xmajorgrids,
ymin=-0.5,
ymax=1,
ytick={-0.5,0,0.5,1},
yticklabels={\empty},
ymajorgrids,
yticklabel style={overlay} %
]
\addplot [color=black,dashed]
  table[row sep=crcr]{figures/coherence-measured-rir-2-1.tsv};

\addplot [color=blue,dash pattern=on 1pt off 3pt on 3pt off 3pt]
  table[row sep=crcr]{figures/coherence-measured-rir-2-2.tsv};

\addplot [color=red,solid]
  table[row sep=crcr]{figures/coherence-measured-rir-2-3.tsv};

\end{axis}

  \path
    ([shift={(-0,0)}]current bounding box.south west)
    ([shift={( 0, 1mm )}]current bounding box.north east);

\end{tikzpicture}%
}
	\hfil\\
	\subfloat[Medium room 1]{\hspace{-11.2mm}%
\begin{tikzpicture}

\begin{axis}[%
width=\figurewidth,
height=\figureheight,
scale only axis,
xmin=0,
xmax=8,
xlabel={$f \mathrm{[kHz]}$},
xmajorgrids,
ymin=-0.5,
ymax=1,
ylabel={$\Re\{\Gamma\}$},
ymajorgrids,
yticklabel style={overlay} %
]
\addplot [color=black,dashed]
  table[row sep=crcr]{figures/coherence-measured-rir-3-1.tsv};

\addplot [color=blue,dash pattern=on 1pt off 3pt on 3pt off 3pt]
  table[row sep=crcr]{figures/coherence-measured-rir-3-2.tsv};

\addplot [color=red,solid]
  table[row sep=crcr]{figures/coherence-measured-rir-3-3.tsv};

\end{axis}

  \path
    ([shift={(-0,0)}]current bounding box.south west)
    ([shift={( 0, 1mm )}]current bounding box.north east);

\end{tikzpicture}%
}
	\subfloat[Medium room 2]{%
\begin{tikzpicture}

\begin{axis}[%
width=\figurewidth,
height=\figureheight,
scale only axis,
xmin=0,
xmax=8,
xlabel={$f \mathrm{[kHz]}$},
xmajorgrids,
ymin=-0.5,
ymax=1,
ytick={-0.5,0,0.5,1},
yticklabels={\empty},
ymajorgrids,
yticklabel style={overlay} %
]
\addplot [color=black,dashed]
  table[row sep=crcr]{figures/coherence-measured-rir-4-1.tsv};

\addplot [color=blue,dash pattern=on 1pt off 3pt on 3pt off 3pt]
  table[row sep=crcr]{figures/coherence-measured-rir-4-2.tsv};

\addplot [color=red,solid]
  table[row sep=crcr]{figures/coherence-measured-rir-4-3.tsv};

\end{axis}

  \path
    ([shift={(-0,0)}]current bounding box.south west)
    ([shift={( 0, 1mm )}]current bounding box.north east);

\end{tikzpicture}%
}
	\hfil\\
	\subfloat[Large room 1]{\hspace{-11.2mm}%
\begin{tikzpicture}

\begin{axis}[%
width=\figurewidth,
height=\figureheight,
scale only axis,
xmin=0,
xmax=8,
xlabel={$f \mathrm{[kHz]}$},
xmajorgrids,
ymin=-0.5,
ymax=1,
ylabel={$\Re\{\Gamma\}$},
ymajorgrids,
yticklabel style={overlay} %
]
\addplot [color=black,dashed]
  table[row sep=crcr]{figures/coherence-measured-rir-5-1.tsv};

\addplot [color=blue,dash pattern=on 1pt off 3pt on 3pt off 3pt]
  table[row sep=crcr]{figures/coherence-measured-rir-5-2.tsv};

\addplot [color=red,solid]
  table[row sep=crcr]{figures/coherence-measured-rir-5-3.tsv};

\end{axis}

  \path
    ([shift={(-0,0)}]current bounding box.south west)
    ([shift={( 0, 1mm )}]current bounding box.north east);

\end{tikzpicture}%
}
	\subfloat[Large room 2]{%
\begin{tikzpicture}

\begin{axis}[%
width=\figurewidth,
height=\figureheight,
scale only axis,
xmin=0,
xmax=8,
xlabel={$f \mathrm{[kHz]}$},
xmajorgrids,
ymin=-0.5,
ymax=1,
ytick={-0.5,0,0.5,1},
yticklabels={\empty},
ymajorgrids,
yticklabel style={overlay} %
]
\addplot [color=black,dashed]
  table[row sep=crcr]{figures/coherence-measured-rir-6-1.tsv};

\addplot [color=blue,dash pattern=on 1pt off 3pt on 3pt off 3pt]
  table[row sep=crcr]{figures/coherence-measured-rir-6-2.tsv};

\addplot [color=red,solid]
  table[row sep=crcr]{figures/coherence-measured-rir-6-3.tsv};

\end{axis}

  \path
    ([shift={(-0,0)}]current bounding box.south west)
    ([shift={( 0, 1mm )}]current bounding box.north east);

\end{tikzpicture}%
}
	\vspace{2mm}
	\hfil\\
\begin{tikzpicture}
    \begin{customlegend}[legend entries={3D isotropic,2D isotropic,measured},legend columns=3,legend style={align=left,/tikz/every even column/.append style={column sep=5pt}},legend cell align={left}]
    \addlegendimage{color=black,dashed}
    \addlegendimage{color=blue,dash pattern=on 1pt off 3pt on 3pt off 3pt}
    \addlegendimage{color=red,solid}
    \end{customlegend}
\end{tikzpicture}
     \caption{Spatial coherence estimated from the reverberation tail of measured RIRs from the REVERB challenge, averaged over 7 microphone pairs with spacing $d=8\,\mathrm{cm}$.}
    \label{fig:coherence-rooms}
\end{figure}
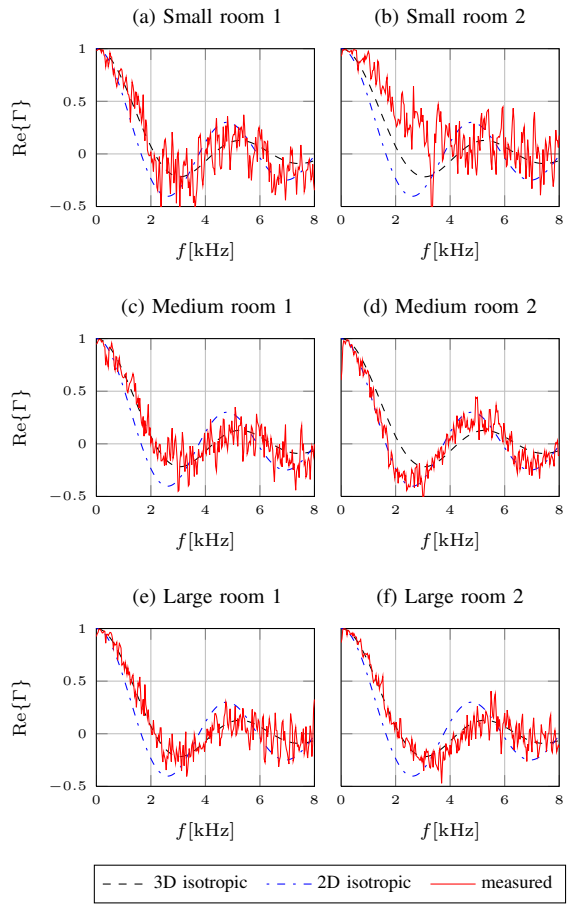

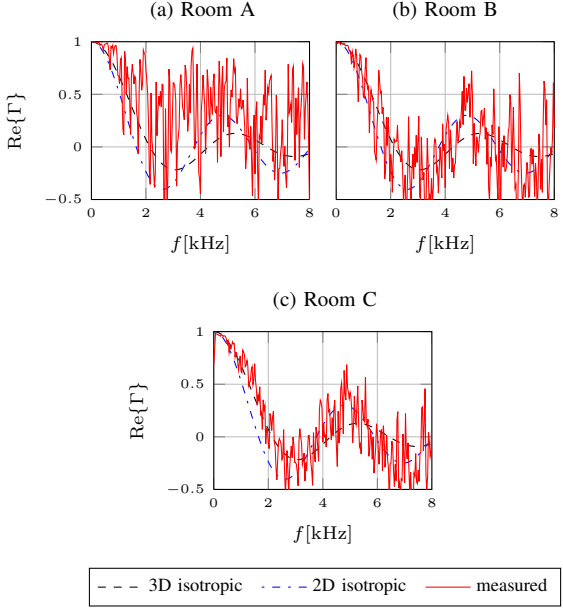
\begin{figure}[t]
    \centering
	\setlength\figureheight{2.1cm}
	\setlength\figurewidth{2.9cm}
\pgfplotsset{
title style={font=\footnotesize},
tick label style={font=\tiny},
label style={font=\scriptsize},
legend style={font=\scriptsize,row sep=-1mm},
}
\subfloat[Room A]{\hspace{-11.2mm}%
\begin{tikzpicture}
\begin{axis}[%
width=\figurewidth,
height=\figureheight,
scale only axis,
xmin=0,
xmax=8,
xlabel={$f \mathrm{[kHz]}$},
xmajorgrids,
ymin=-0.5,
ymax=1,
ylabel={$\Re\{\Gamma\}$},
ymajorgrids,
yticklabel style={overlay} %
]
\addplot [color=black,dashed]
  table[row sep=crcr]{figures/coherence-measured-abc-rir-1-1.tsv};

\addplot [color=blue,dash pattern=on 1pt off 3pt on 3pt off 3pt]
  table[row sep=crcr]{figures/coherence-measured-abc-rir-1-2.tsv};

\addplot [color=red,solid]
  table[row sep=crcr]{figures/coherence-measured-abc-rir-1-3.tsv};

\end{axis}
  \path
    ([shift={(-0,0)}]current bounding box.south west)
    ([shift={( 0, 1mm )}]current bounding box.north east);
\end{tikzpicture}%
}
\subfloat[Room B]{%
\begin{tikzpicture}
\begin{axis}[%
width=\figurewidth,
height=\figureheight,
scale only axis,
xmin=0,
xmax=8,
xlabel={$f \mathrm{[kHz]}$},
xmajorgrids,
ymin=-0.5,
ymax=1,
ytick={-0.5,0,0.5,1},
yticklabels={\empty},
ymajorgrids,
yticklabel style={overlay} %
]
\addplot [color=black,dashed]
  table[row sep=crcr]{figures/coherence-measured-abc-rir-2-1.tsv};

\addplot [color=blue,dash pattern=on 1pt off 3pt on 3pt off 3pt]
  table[row sep=crcr]{figures/coherence-measured-abc-rir-2-2.tsv};

\addplot [color=red,solid]
  table[row sep=crcr]{figures/coherence-measured-abc-rir-2-3.tsv};

\end{axis}
  \path
    ([shift={(-0,0)}]current bounding box.south west)
    ([shift={( 0, 1mm )}]current bounding box.north east);
\end{tikzpicture}%
}\hfil\\
\subfloat[Room C]{\hfil\hspace{-11.2mm}%
\begin{tikzpicture}
\begin{axis}[%
width=\figurewidth,
height=\figureheight,
scale only axis,
xmin=0,
xmax=8,
xlabel={$f \mathrm{[kHz]}$},
xmajorgrids,
ymin=-0.5,
ymax=1,
ylabel={$\Re\{\Gamma\}$},
ymajorgrids,
yticklabel style={overlay} %
]
\addplot [color=black,dashed]
  table[row sep=crcr]{figures/coherence-measured-abc-rir-3-1.tsv};

\addplot [color=blue,dash pattern=on 1pt off 3pt on 3pt off 3pt]
  table[row sep=crcr]{figures/coherence-measured-abc-rir-3-2.tsv};

\addplot [color=red,solid]
  table[row sep=crcr]{figures/coherence-measured-abc-rir-3-3.tsv};

\end{axis}
  \path
    ([shift={(-0,0)}]current bounding box.south west)
    ([shift={( 0, 1mm )}]current bounding box.north east);
\end{tikzpicture}%
}
\vspace{2mm}\hfil\\%
\begin{tikzpicture}
    \begin{customlegend}[legend entries={3D isotropic,2D isotropic,measured},legend columns=3,legend style={align=left,/tikz/every even column/.append style={column sep=5pt}},legend cell align={left}]
    \addlegendimage{color=black,dashed}
    \addlegendimage{color=blue,dash pattern=on 1pt off 3pt on 3pt off 3pt}
    \addlegendimage{color=red,solid}
    \end{customlegend}
\end{tikzpicture}
     \caption{Spatial coherence estimated from the reverberation tail of measured RIRs in rooms A, B, C, one microphone pair with spacing $d=8\,\mathrm{cm}$.}
    \label{fig:coherence-measured-rooms}
\end{figure}

First, RIRs are generated using the image method \cite{allen_image_1979,peterson_simulating_1986}. In the simulations, a uniform linear array (inter-microphone spacing $d=8\,\mathrm{cm}$) is placed horizontally in the center of rectangular rooms with varying dimensions and reflectivities. The image source order is chosen sufficiently high to include all reflections within $60\,\mathrm{dB}$ of the main peak. In order to reduce the variance of the estimate for a better visualization, the coherence is also spatially averaged over the estimates from 7 microphone pairs \cite{jacobsen_coherence_2000}. Fig.~\ref{fig:coherence-simulated-rooms} shows plots of the real part of the resulting coherence, for a large room ($15\times18\times10\,$m, left) and a small room ($4\times3\times2.5\,$m, right); for both rooms, three configurations for the surface reflectivity $\beta$ are used: equally high reflectivity for all surfaces ($\beta=0.9$), highly absorbing floor and ceiling ($\beta_\text{Walls}=0.9$, $\beta_\text{Floor, Ceil} = 0.1$), and moderately absorbing walls ($\beta_\text{Walls}=0.5$, $\beta_\text{Floor, Ceil} = 0.9$). The results in Fig.~\ref{fig:coherence-simulated-rooms} confirm the assumptions on the coherence properties of reverberation that were made in Section~\ref{sec:reverberation-coherence}: for equal reflectivity of all surfaces, the coherence closely matches the coherence of the diffuse sound field. If floor and ceiling are highly absorbing, the model of a 2D isotropic sound field is appropriate. If instead the walls are more absorbing than floor and ceiling, the coherence is significantly higher than the diffuse coherence, since the dominating vertically propagating components are strongly correlated between the horizontally spaced microphones. Also, the variance of the coherence estimate is visibly lower in the larger room. 

Fig.~\ref{fig:coherence-rooms} shows the reverberation coherence estimates obtained from the RIRs of the REVERB challenge database, estimated in the same way as for the simulated RIRs. The coherence estimates are obtained for 7 pairs of neighboring microphones from the circular array and averaged. Most rooms match the diffuse model quite well, with two exceptions. In SR2, the coherence is higher than expected from the diffuse model, which can be explained by the presence of absorbing curtains on all four walls. In MR2, the coherence however almost perfectly matches the 2D isotropic model, since in this room, walls are more reflective than floor and ceiling. Also, it can again be observed that the variance of the coherence estimate is lower for rooms with a longer reverberation time.

Fig.~\ref{fig:coherence-measured-rooms} shows the results for one position in the rooms A, B and C. The coherence estimate is here computed just from one pair of microphones, therefore the variance is significantly higher. The diffuse model is a good fit for rooms B and C, where all surfaces are highly reflective. In room A, the coherence is similar to the simulated case of partially absorbing walls, which is due to the presence of partially closed curtains on the walls of the room.

Concluding the analysis of the spatial properties, it can be stated that, for microphones located in the same horizontal plane, the spatial coherence of reverberation in real rooms typically lies between the coherence of diffuse and 2D isotropic noise, with some exceptions where the coherence is increased due to dominant vertical reflections. The diffuse model is a good fit for most rooms, unless there are large differences in the reflectivity of the room surfaces.
Finally, it is noteworthy that the image source model with sufficient order can reproduce the spatial characteristics of late reverberation which are observed in real rooms.

\subsection{CDR Estimation for Reverberant Speech}
\label{sec:cdr_for_reverberant_speech}
In Section~\ref{sec:signal-model}, a reverberant speech signal is modeled as consisting of a directional and a diffuse component, which are mutually uncorrelated. In practice, the reverberant sound field consists of the direct path, several spatially distinct early reflections, and the reverberation component, all of which are not perfectly uncorrelated, due to the non-zero length of the observation window and the temporal correlation of speech signals. In the previous section, it was shown that the model of a diffuse sound field is appropriate for the reverberation component. In the following, it is investigated whether the simplified model of a mixture of uncorrelated directional and diffuse sound fields can be applied to real reverberant speech signals, i.e., whether the CDR estimate can be used as a practical measure for the time- and frequency-dependent ratio between desired and undesired signal components, as it is required for speech enhancement. We now consider the desired signal components to be the direct path plus the reflections arriving within $T_\mathrm{e}=50\,\mathrm{ms}$ after the direct path, and the undesired components to be the energy caused by the reverberation tail of the RIR. This is motivated by the well-known effect that early reflections are beneficial both for speech intelligibility \cite{bradley_importance_2003} and ASR accuracy \cite{sehr_towards_2010}, and should therefore be considered part of the desired signal. In other words, the relevant SNR to be estimated for the application to signal enhancement is the early-to-late power ratio $\ELR_{50\,\mathrm{ms}}(l,f)$ (see Appendix).

\begin{figure}[tb]
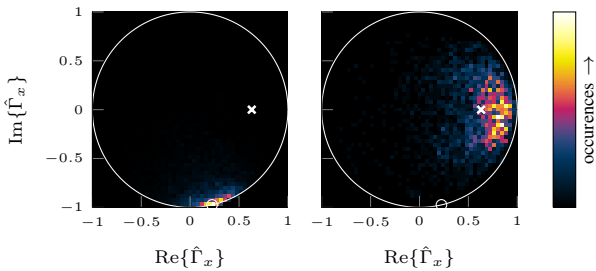

    \centering
        \setlength\figureheight{2.6cm}
	\setlength\figurewidth{2.6cm}
\pgfplotsset{
tick label style={font=\tiny},
label style={font=\scriptsize},
legend style={font=\footnotesize},
title style={font=\footnotesize,text height=1em},
}
%
    \caption{Histogram of complex coherence values $\hat\Gamma_x$ measured from a reverberant speech signal, for time-frequency bins with $\ELR_{50\,\mathrm{ms}} > 10\,\dB$ (left) and $< -10\,\dB$ (right). Room B, $l=2\,\mathrm{m}$, $d=8\,\mathrm{cm}$, $\theta=60^\circ$, $f=1\,\mathrm{kHz}$). Theoretical signal coherence $\Gamma_s$ computed from measured TDOA and diffuse noise coherence $\Gamma_n$ are marked by $\mathbf{\circ}$ and $\mathbf{\times}$, respectively.}
    \label{fig:histogram}
\end{figure}

To exemplarily illustrate the relationship between the (non-stationary) early-to-late power ratio and the short-time coherence estimate, the time-frequency bins of a reverberant speech signal are first classified according to the instantaneous $\ELR_{50\,\mathrm{ms}}$ into low-reverberant and highly reverberant, and the corresponding distribution of the short-time estimates of the complex coherence is visualized as a histogram. Fig.~\ref{fig:histogram} shows the two-dimensional histograms of the complex coherence of bins with $\ELR > 10\,\dB$ (left) and $\ELR < -10\,\dB$ (right) around $f=1\,\mathrm{kHz}$. The coherence of the low-reverberant bins matches the coherence of a single plane wave quite well, although the signal contains contributions from early reflections in addition to the direct path. The phase has a slight spread, caused by early reflections; this has to be tolerated by the CDR estimator. The coherence of the highly reverberant bins, which should lie close to the diffuse model coherence, has a considerably higher spread and is not exactly centered around the model. This indicates that, while the simplified model seems to be reasonable, errors are non-negligible, and the differences in the realizations of the unbiased estimators, which affect only the behavior for values deviating from the ideal model, are likely to have a significant impact on estimation performance.

For the comparison of the estimation performance of the different estimators, it is convenient to transform the true and estimated CDR into the true and estimated diffuseness $D=[\CDR+1]^{-1}$ and $\hat D=[\widehat\CDR+1]^{-1}$, respectively, due to the diffuseness being bounded between 0 and 1, and to evaluate the mean squared error $\widehat{\MSE} = \E\{|D-\hat D|^2\}$. For this evaluation, the true CDR is again approximated by the ELR ($CDR\approx\ELR_{50\,\mathrm{ms}}$), and the expectation is approximated by averaging over time and frequency. The coherence models $\tilde\Gamma_s$ and $\tilde\Gamma_n$ for the estimators are based on the measured TDOA and the diffuse coherence assumption, respectively. Table~\ref{table:estimationerror} shows the MSE for the different estimators, averaged over all source positions in the respective room. The estimator $\CDRthiergartone$ has a relatively high estimation error, due to the high sensitivity of this estimator towards phase variation of the coherence. The estimator $\CDRpropone$ shows a slightly reduced estimation error compared to the biased estimator $\CDRjeub$, while the variant $\CDRproptwo$ further reduces the error. Among the DOA-independent estimators, the proposed unbiased version leads to an error reduction as well, while the noise coherence-independent variant $\CDRpropfour$ has the overall second-highest error, due to the difficulties in cases where the phase of the coherence is close to zero.

\begin{table}[t]
    \centering
    \caption{Estimation error of different CDR estimators.}
    \includegraphics[width=\columnwidth]{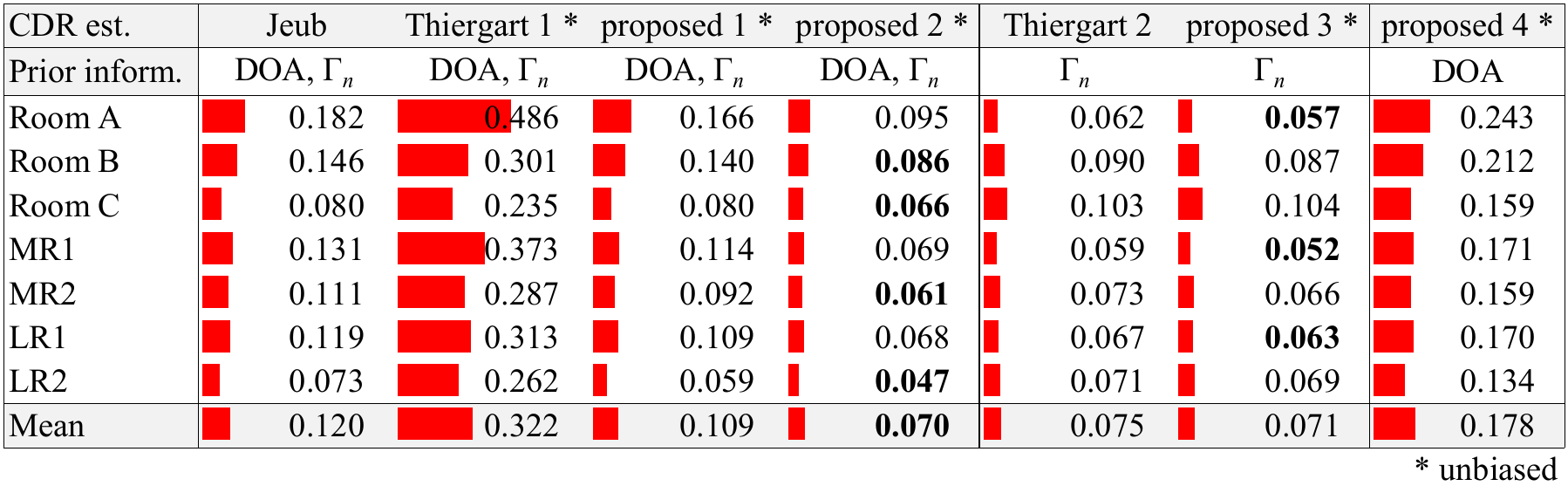}
    \label{table:estimationerror}
\end{table}

\subsection{Dereverberation Performance}
\label{sec:asr}

In the following, the signal enhancement system described in Section~\ref{sec:ReverberationSuppression} is evaluated for the application to dereverberation. For all of the following results, two-channel signals are processed by first applying spatial magnitude averaging as described by (\ref{eq:psdavg}), and then applying a postfilter based on the different CDR estimators, or one of several other dereverberation methods used for comparison.

\subsubsection{Measures and Evaluation Method}

To quantify the amount of reverberation in the unprocessed and processed signals, the time- and frequency-averaged early-to-late power ratio $\ELR_{50\,\mathrm{ms}}$ is evaluated (see Appendix). The amount of signal distortion caused by the postfilter is quantified by the frequency-weighted segmental signal-to-distortion ratio (fwSegSDR), which we define as the fwSegSNR \cite{hu_evaluation_2008} computed for the postfiltered early signal component (i.e., the signal convolved with the first $50\,\mathrm{ms}$ of the RIR), with the unprocessed early signal component $Y_e$ as the reference:
\begin{flalign}
\fwSegSDR=\fwSegSNR(Y_{e}(\frameix,f), G(\frameix,f) Y_{e}(\frameix,f))
\end{flalign}

The overall quality of the processed signals, including both the effects of reverberation reduction and undesired speech distortion, is evaluated using the recognition rate of an automatic speech recognizer. The ASR engine PocketSphinx \cite{huggins-daines_pocketsphinx:_2006} is used with an acoustic model trained on clean speech from the GRID corpus \cite{cooke_audio-visual_2006}, using MFCC+$\Delta$+$\Delta\Delta$ features. Cepstral mean normalization is used for the equalization of the effect of early reverberation \cite{furui_cepstral_1981}. For the computation of the recognition rate, only the letter and the number in the utterance are evaluated, as in the CHiME challenge \cite{christensen_chime_2010}. Furthermore, two signal-based measures for the overall speech quality are evaluated, which were shown to be significantly correlated to the perceived amount of reverberation \cite{goetze_study_2014}: PESQ \cite{rix_perceptual_2001} and the frequency-weighted segmental signal-to-noise ratio (fwSegSNR) \cite{hu_evaluation_2008}. We use the wideband version of PESQ and give values in the MOS-LQO scale. For both PESQ and the fwSegSNR, the clean speech signal is used as reference.

CDR-based dereverberation is evaluated with all estimators discussed in this paper. In addition to the CDR-based methods, two heuristic coherence-based postfiltering methods are evaluated: a version of Allen's method \cite{allen_multimicrophone_1977}, where the magnitude of the coherence is used as a spectral gain and applied to the spatially preprocessed signal, and the coherence-to-gain-mapping proposed by Westermann et al. \cite{westermann_binaural_2013}, which depends on a histogram of the magnitude squared coherence. Also evaluated is the exponential decay model by Lebart et al. \cite{lebart_new_2001}, using the true reverberation times measured from the RIRs, which in practice would have to be estimated blindly from the reverberant signals \cite{schuldt_decay_2014}. For the method of Lebart and the CDR-based methods, spectral magnitude subtraction according to (\ref{eq:Magnitude_Subtraction}) is applied, with $G_\mathrm{min}=0.1$. The suppression parameter $\mu$ is set to $1.3$, which yields close to optimum recognition rates for all except Lebart's method (see the comment in the following section).
Ideal TDOA knowledge is assumed for the CDR estimators which require a TDOA estimate $\widehat{\Delta t}$, i.e., $\widehat{\Delta t}=\Delta t$. The dereverberation methods are evaluated for the rooms A, B, C, MR1/2 and LR1/2. In SR1/2, the very low amount of reverberation ($T_{60} < 0.3\,\mathrm{s}$) did not lead to a significantly lower recognition rate compared to clean speech, therefore these rooms are not included in the evaluation.
For each room and source position, 500 GRID utterances are convolved with the measured two-channel RIRs (in the case of the REVERB challenge RIRs, two neighboring microphones are selected from the circular array), and then processed by the dereverberation methods.

\subsubsection{Results}

Table~\ref{table:results} summarizes the resulting performance measurements, averaged over all source positions in each room. The first column shows the results for the unprocessed microphone signals. The spatial magnitude averaging leads to a small but consistent improvement in all performance measures, as seen in the second column.

\begin{table*}[!t]
    \centering
    \caption{Performance measures, averaged over all source positions in each room. First column: unprocessed microphone signal, second column: spatially averaged magnitudes without postfiltering, remaining columns: different postfilters.}
    \label{table:results}
    \includegraphics[width=\textwidth]{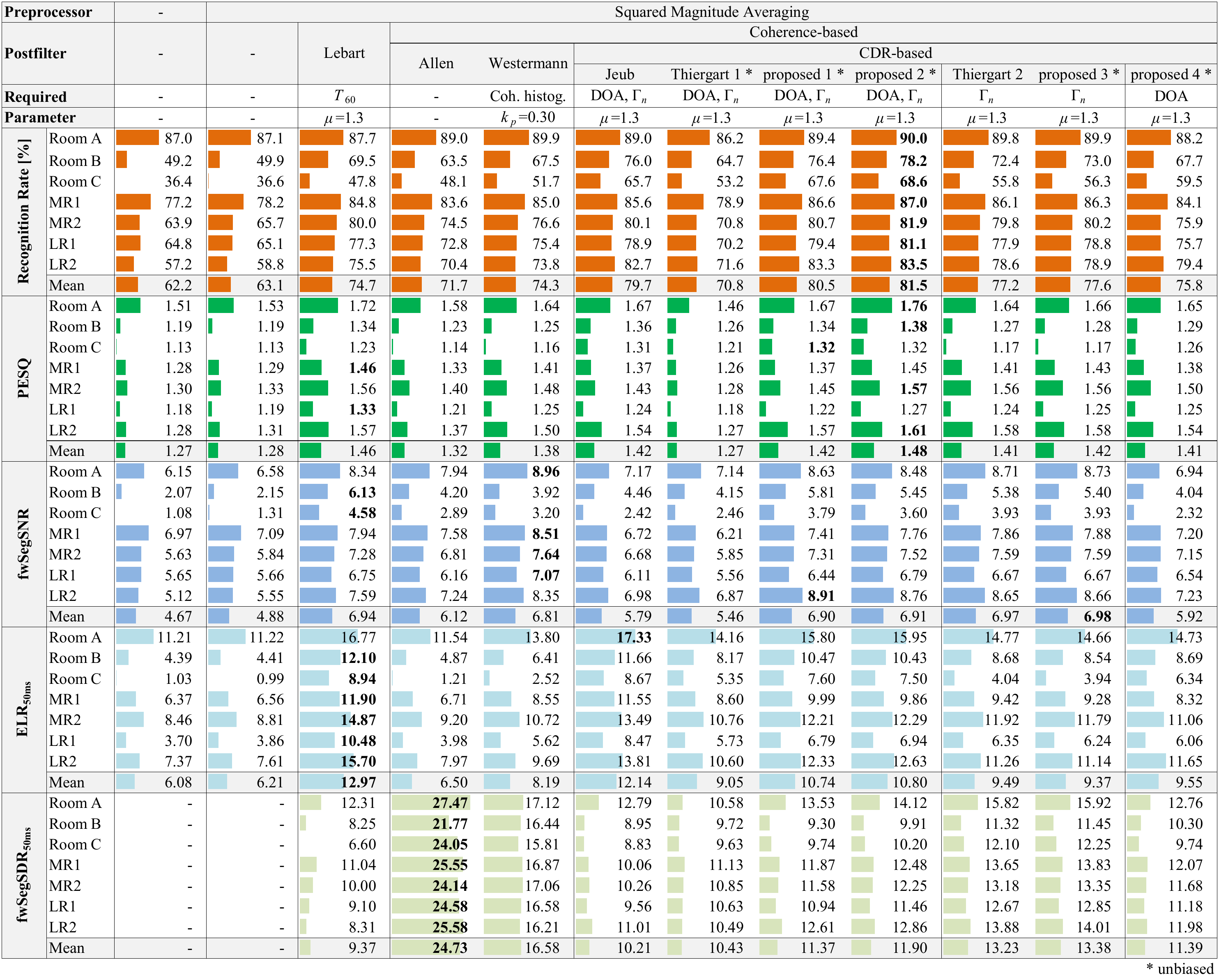}
\end{table*}

Postfiltering using the CDR estimator $\CDRproptwo$ leads to the highest recognition rate among all methods across all evaluated rooms, as well as to the highest average PESQ score. Comparing the CDR-based methods, the following observations can be made: both for the DOA-dependent and DOA-independent estimators, all measures reflect the slight advantage of the respective unbiased variant ($\CDRpropone$ and $\CDRpropthree$, respectively) over the biased estimators. For the DOA-dependent estimator, the variant $\CDRproptwo$ further improves the result over the first proposed unbiased estimator, due to the different behavior of this estimator for coherence values which deviate from the ideal coherence model. The significant improvement suggests that further improvement may be possible by modeling these deviations statistically and explicitly optimizing the estimator for this model. Remarkable are the results of the DOA-independent estimators: without requiring any knowledge or estimation of source DOA or other parameters of the scenario, the CDR-based postfilter can significantly increase the overall signal quality according to all evaluated measures.

Compared to CDR-based dereverberation, the methods by Allen and Westermann yield a low ELR improvement, and at the same time a higher signal-to-distortion ratio. The overall improvement in recognition rate and PESQ is relatively low for both, while Westermann's method shows good results for the fwSegSNR. The discrepancies between these measures can be explained by the different tradeoffs between reverberation suppression and signal distortion, which have different effects on the evaluated quality measures. Apparently, Allen's and Westermann's methods apply a lower overall amount of suppression, which benefits the fwSegSNR measure, but has a small effect on ASR recognition rate and PESQ.

It is noticeable that Lebart's method yields the highest ELR, but at the same time the worst signal-to-distortion ratio; this indicates that reverberation is overestimated, and consequently too much suppression is applied, possibly due to mismatch between the exponential decay assumption and the early part of the impulse responses \cite{habets_late_2009}. Reducing the suppression gain to the optimum value $\mu=0.6$ to counter overestimation increases the mean recognition rate to $77.4\,\%$.

\begin{figure}[tb]
    \centering
	\setlength\figureheight{2.3cm}
	\setlength\figurewidth{2.3cm}
\pgfplotsset{
title style={font=\footnotesize},
tick label style={font=\tiny},
label style={font=\scriptsize},
legend style={font=\scriptsize},
/tikz/mark size=1.5pt
}
\subfloat[CDR estimator $\widehat{\CDR}_\text{prop2}$]{\hspace{-1mm}%
\begin{tikzpicture}
\begin{groupplot}[%
group style={
	columns=3,
	rows=1,
	xlabels at=edge bottom,
	ylabels at=edge left,
	horizontal sep=6mm,
	vertical sep=16mm
},
height=\figureheight,
width=\figurewidth,
scale only axis,
y label style={at={(-0.12,0.5)}},
]

\nextgroupplot[%
scale only axis,
xmin=-90,
xmax=90,
xlabel={DOA $\theta$},
xmajorgrids,
ymin=70,
ymax=95,
xtick={-45,0,45},
ylabel={recognition rate [\%]},
ymajorgrids,
title={Room A}
]
\addplot [color=red,dotted,mark=o,mark options={solid}]
  table[row sep=crcr]{figures/wer-angles-distances-cdr_proposed_robust_unbiased-roomA-1.tsv};

\addplot [color=blue,dotted,mark=x,mark options={solid}]
  table[row sep=crcr]{figures/wer-angles-distances-cdr_proposed_robust_unbiased-roomA-2.tsv};

\addplot [color=black,dotted,mark=diamond,mark options={solid}]
  table[row sep=crcr]{figures/wer-angles-distances-cdr_proposed_robust_unbiased-roomA-3.tsv};

\addplot [color=red,solid,mark=o,mark options={solid}]
  table[row sep=crcr]{figures/wer-angles-distances-cdr_proposed_robust_unbiased-roomA-4.tsv};

\addplot [color=blue,solid,mark=x,mark options={solid}]
  table[row sep=crcr]{figures/wer-angles-distances-cdr_proposed_robust_unbiased-roomA-5.tsv};

\addplot [color=black,solid,mark=diamond,mark options={solid}]
  table[row sep=crcr]{figures/wer-angles-distances-cdr_proposed_robust_unbiased-roomA-6.tsv};

\nextgroupplot[%
scale only axis,
xmin=-90,
xmax=90,
xlabel={DOA $\theta$},
xmajorgrids,
xtick={-45,0,45},
ymin=20,
ymax=100,
ymajorgrids,
title={Room B}
]
\addplot [color=red,dotted,mark=o,mark options={solid}]
  table[row sep=crcr]{figures/wer-angles-distances-cdr_proposed_robust_unbiased-roomB-1.tsv};

\addplot [color=blue,dotted,mark=x,mark options={solid}]
  table[row sep=crcr]{figures/wer-angles-distances-cdr_proposed_robust_unbiased-roomB-2.tsv};

\addplot [color=black,dotted,mark=diamond,mark options={solid}]
  table[row sep=crcr]{figures/wer-angles-distances-cdr_proposed_robust_unbiased-roomB-3.tsv};

\addplot [color=red,solid,mark=o,mark options={solid}]
  table[row sep=crcr]{figures/wer-angles-distances-cdr_proposed_robust_unbiased-roomB-4.tsv};

\addplot [color=blue,solid,mark=x,mark options={solid}]
  table[row sep=crcr]{figures/wer-angles-distances-cdr_proposed_robust_unbiased-roomB-5.tsv};

\addplot [color=black,solid,mark=diamond,mark options={solid}]
  table[row sep=crcr]{figures/wer-angles-distances-cdr_proposed_robust_unbiased-roomB-6.tsv};

\nextgroupplot[%
scale only axis,
xmin=-90,
xmax=90,
xlabel={DOA $\theta$},
xmajorgrids,
xtick={-45,0,45},
ymin=0,
ymax=100,
ymajorgrids,
title={Room C},
]
\addplot [color=red,dotted,mark=o,mark options={solid}]
  table[row sep=crcr]{figures/wer-angles-distances-cdr_proposed_robust_unbiased-roomC-1.tsv};
\addplot [color=blue,dotted,mark=x,mark options={solid}]
  table[row sep=crcr]{figures/wer-angles-distances-cdr_proposed_robust_unbiased-roomC-2.tsv};

\addplot [color=black,dotted,mark=diamond,mark options={solid}]
  table[row sep=crcr]{figures/wer-angles-distances-cdr_proposed_robust_unbiased-roomC-3.tsv};

\addplot [color=red,solid,mark=o,mark options={solid}]
  table[row sep=crcr]{figures/wer-angles-distances-cdr_proposed_robust_unbiased-roomC-4.tsv};

\addplot [color=blue,solid,mark=x,mark options={solid}]
  table[row sep=crcr]{figures/wer-angles-distances-cdr_proposed_robust_unbiased-roomC-5.tsv};

\addplot [color=black,solid,mark=diamond,mark options={solid}]
  table[row sep=crcr]{figures/wer-angles-distances-cdr_proposed_robust_unbiased-roomC-6.tsv};
\end{groupplot}

\end{tikzpicture}%
 }\hfil
\subfloat[CDR estimator $\widehat{\CDR}_\text{prop3}$ (DOA-independent)]{\hspace{-1mm}%
\begin{tikzpicture}
\begin{groupplot}[%
group style={
	columns=3,
	rows=1,
	xlabels at=edge bottom,
	ylabels at=edge left,
	horizontal sep=6mm,
	vertical sep=16mm,
	group name=plots
},
height=\figureheight,
width=\figurewidth,
scale only axis,
y label style={at={(-0.12,0.5)}},
legend to name=grouplegend2,
]

\nextgroupplot[%
scale only axis,
xmin=-90,
xmax=90,
xlabel={DOA $\theta$},
xmajorgrids,
ymin=70,
ymax=95,
xtick={-45,0,45},
ylabel={recognition rate [\%]},
ymajorgrids,
title={Room A}
]
\addplot [color=red,dotted,mark=o,mark options={solid}]
  table[row sep=crcr]{figures/wer-angles-distances-cdr_proposed_nodoa-roomA-1.tsv};

\addplot [color=blue,dotted,mark=x,mark options={solid}]
  table[row sep=crcr]{figures/wer-angles-distances-cdr_proposed_nodoa-roomA-2.tsv};

\addplot [color=black,dotted,mark=diamond,mark options={solid}]
  table[row sep=crcr]{figures/wer-angles-distances-cdr_proposed_nodoa-roomA-3.tsv};

\addplot [color=red,solid,mark=o,mark options={solid}]
  table[row sep=crcr]{figures/wer-angles-distances-cdr_proposed_nodoa-roomA-4.tsv};

\addplot [color=blue,solid,mark=x,mark options={solid}]
  table[row sep=crcr]{figures/wer-angles-distances-cdr_proposed_nodoa-roomA-5.tsv};

\addplot [color=black,solid,mark=diamond,mark options={solid}]
  table[row sep=crcr]{figures/wer-angles-distances-cdr_proposed_nodoa-roomA-6.tsv};

\nextgroupplot[%
scale only axis,
xmin=-90,
xmax=90,
xlabel={DOA $\theta$},
xmajorgrids,
xtick={-45,0,45},
ymin=20,
ymax=100,
ymajorgrids,
title={Room B}
]
\addplot [color=red,dotted,mark=o,mark options={solid}]
  table[row sep=crcr]{figures/wer-angles-distances-cdr_proposed_nodoa-roomB-1.tsv};

\addplot [color=blue,dotted,mark=x,mark options={solid}]
  table[row sep=crcr]{figures/wer-angles-distances-cdr_proposed_nodoa-roomB-2.tsv};

\addplot [color=black,dotted,mark=diamond,mark options={solid}]
  table[row sep=crcr]{figures/wer-angles-distances-cdr_proposed_nodoa-roomB-3.tsv};

\addplot [color=red,solid,mark=o,mark options={solid}]
  table[row sep=crcr]{figures/wer-angles-distances-cdr_proposed_nodoa-roomB-4.tsv};

\addplot [color=blue,solid,mark=x,mark options={solid}]
  table[row sep=crcr]{figures/wer-angles-distances-cdr_proposed_nodoa-roomB-5.tsv};

\addplot [color=black,solid,mark=diamond,mark options={solid}]
  table[row sep=crcr]{figures/wer-angles-distances-cdr_proposed_nodoa-roomB-6.tsv};

\nextgroupplot[%
scale only axis,
xmin=-90,
xmax=90,
xlabel={DOA $\theta$},
xmajorgrids,
xtick={-45,0,45},
ymin=0,
ymax=100,
ymajorgrids,
title={Room C},
]
\addplot [color=red,dotted,mark=o,mark options={solid}]
  table[row sep=crcr]{figures/wer-angles-distances-cdr_proposed_nodoa-roomC-1.tsv};

\addplot [color=blue,dotted,mark=x,mark options={solid}]
  table[row sep=crcr]{figures/wer-angles-distances-cdr_proposed_nodoa-roomC-2.tsv};

\addplot [color=black,dotted,mark=diamond,mark options={solid}]
  table[row sep=crcr]{figures/wer-angles-distances-cdr_proposed_nodoa-roomC-3.tsv};

\addplot [color=red,solid,mark=o,mark options={solid}]
  table[row sep=crcr]{figures/wer-angles-distances-cdr_proposed_nodoa-roomC-4.tsv};

\addplot [color=blue,solid,mark=x,mark options={solid}]
  table[row sep=crcr]{figures/wer-angles-distances-cdr_proposed_nodoa-roomC-5.tsv};

\addplot [color=black,solid,mark=diamond,mark options={solid}]
  table[row sep=crcr]{figures/wer-angles-distances-cdr_proposed_nodoa-roomC-6.tsv};
\end{groupplot}

\end{tikzpicture}%
 }\hfil
\subfloat[CDR estimator $\widehat{\CDR}_\text{prop4}$ (no noise coherence model)]{\hspace{-1mm}%
\begin{tikzpicture}
\begin{groupplot}[%
group style={
	columns=3,
	rows=1,
	xlabels at=edge bottom,
	ylabels at=edge left,
	horizontal sep=6mm,
	vertical sep=16mm,
	group name=plots
},
height=\figureheight,
width=\figurewidth,
scale only axis,
y label style={at={(-0.12,0.5)}},
legend to name=grouplegend12312332,
legend columns=3,
legend style={align=left,/tikz/every even column/.append style={column sep=5pt}},legend cell align={left}
]

\nextgroupplot[%
scale only axis,
xmin=-90,
xmax=90,
xlabel={DOA $\theta$},
xmajorgrids,
ymin=70,
ymax=95,
xtick={-45,0,45},
ylabel={recognition rate [\%]},
ymajorgrids,
title={Room A}
]
\addplot [color=red,dotted,mark=o,mark options={solid}]
  table[row sep=crcr]{figures/wer-angles-distances-cdr_proposed_nodiffuse-roomA-1.tsv};

\addplot [color=blue,dotted,mark=x,mark options={solid}]
  table[row sep=crcr]{figures/wer-angles-distances-cdr_proposed_nodiffuse-roomA-2.tsv};

\addplot [color=black,dotted,mark=diamond,mark options={solid}]
  table[row sep=crcr]{figures/wer-angles-distances-cdr_proposed_nodiffuse-roomA-3.tsv};

\addplot [color=red,solid,mark=o,mark options={solid}]
  table[row sep=crcr]{figures/wer-angles-distances-cdr_proposed_nodiffuse-roomA-4.tsv};

\addplot [color=blue,solid,mark=x,mark options={solid}]
  table[row sep=crcr]{figures/wer-angles-distances-cdr_proposed_nodiffuse-roomA-5.tsv};

\addplot [color=black,solid,mark=diamond,mark options={solid}]
  table[row sep=crcr]{figures/wer-angles-distances-cdr_proposed_nodiffuse-roomA-6.tsv};

\nextgroupplot[%
scale only axis,
xmin=-90,
xmax=90,
xlabel={DOA $\theta$},
xmajorgrids,
xtick={-45,0,45},
ymin=20,
ymax=100,
ymajorgrids,
title={Room B}
]
\addplot [color=red,dotted,mark=o,mark options={solid}]
  table[row sep=crcr]{figures/wer-angles-distances-cdr_proposed_nodiffuse-roomB-1.tsv};

\addplot [color=blue,dotted,mark=x,mark options={solid}]
  table[row sep=crcr]{figures/wer-angles-distances-cdr_proposed_nodiffuse-roomB-2.tsv};

\addplot [color=black,dotted,mark=diamond,mark options={solid}]
  table[row sep=crcr]{figures/wer-angles-distances-cdr_proposed_nodiffuse-roomB-3.tsv};

\addplot [color=red,solid,mark=o,mark options={solid}]
  table[row sep=crcr]{figures/wer-angles-distances-cdr_proposed_nodiffuse-roomB-4.tsv};

\addplot [color=blue,solid,mark=x,mark options={solid}]
  table[row sep=crcr]{figures/wer-angles-distances-cdr_proposed_nodiffuse-roomB-5.tsv};

\addplot [color=black,solid,mark=diamond,mark options={solid}]
  table[row sep=crcr]{figures/wer-angles-distances-cdr_proposed_nodiffuse-roomB-6.tsv};

\nextgroupplot[%
scale only axis,
xmin=-90,
xmax=90,
xlabel={DOA $\theta$},
xmajorgrids,
xtick={-45,0,45},
ymin=0,
ymax=100,
ymajorgrids,
title={Room C},
legend style={draw=black,fill=white,legend cell align=left,at={(1.2,0.5)},anchor=west}
]

\addplot [color=red,solid,mark=o,mark options={solid}]
  table[row sep=crcr]{figures/wer-angles-distances-cdr_proposed_nodiffuse-roomC-4.tsv};
\addlegendentry{1m, processed};

\addplot [color=blue,solid,mark=x,mark options={solid}]
  table[row sep=crcr]{figures/wer-angles-distances-cdr_proposed_nodiffuse-roomC-5.tsv};
\addlegendentry{2m, processed};

\addplot [color=black,solid,mark=diamond,mark options={solid}]
  table[row sep=crcr]{figures/wer-angles-distances-cdr_proposed_nodiffuse-roomC-6.tsv};
\addlegendentry{4m, processed};

\addplot [color=red,dotted,mark=o,mark options={solid}]
  table[row sep=crcr]{figures/wer-angles-distances-cdr_proposed_nodiffuse-roomC-1.tsv};
\addlegendentry{1m, unprocessed};

\addplot [color=blue,dotted,mark=x,mark options={solid}]
  table[row sep=crcr]{figures/wer-angles-distances-cdr_proposed_nodiffuse-roomC-2.tsv};
\addlegendentry{2m, unprocessed};

\addplot [color=black,dotted,mark=diamond,mark options={solid}]
  table[row sep=crcr]{figures/wer-angles-distances-cdr_proposed_nodiffuse-roomC-3.tsv};
\addlegendentry{4m, unprocessed};

\end{groupplot}

\node at (plots c2r1.south) [inner sep=0pt,anchor=north, yshift=-1cm] {\pgfplotslegendfromname{grouplegend12312332}};
 
\end{tikzpicture}%
 }\hfil
    \caption{Average recognition rate for different rooms and source positions ($l=$1,2,4\,m, $\theta=-90\dots90^\circ$), for unprocessed signals and signals processed by spatial magnitude averaging combined with coherence-based postfilters based on different CDR estimators.}
    \label{fig:ASR-angle}
\end{figure}
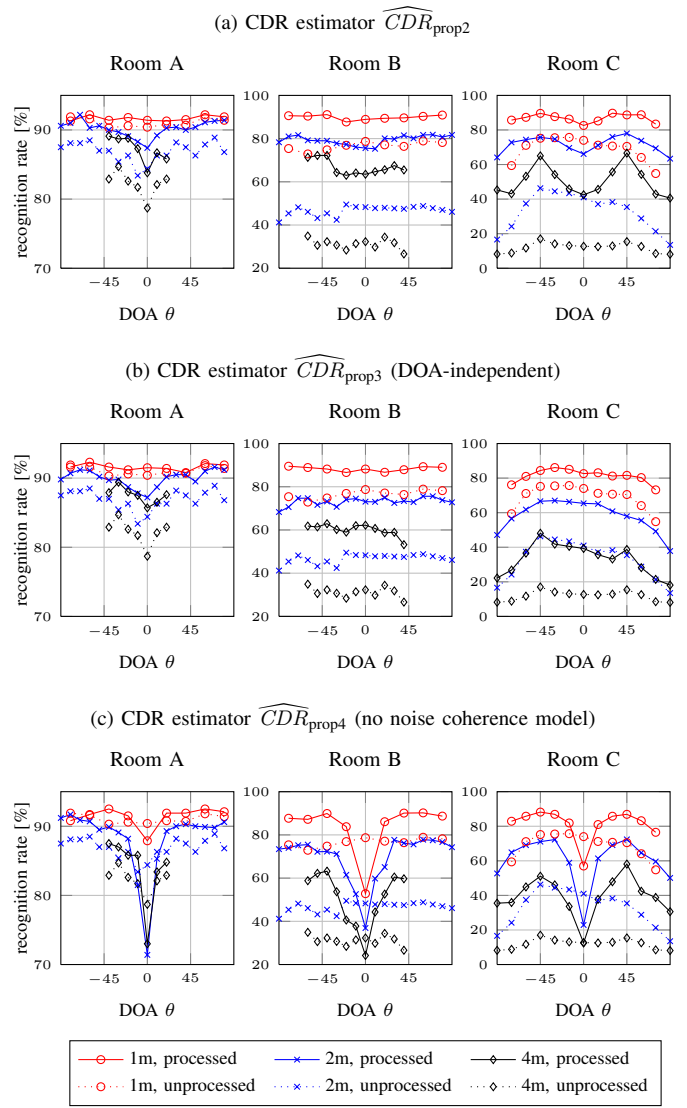

The estimator $\CDRpropfour$, which makes no assumption on the noise coherence, yields on average comparable results to the other estimators, although it can not obtain usable CDR estimates for some of the source positions where the TDOA is close to zero. To gain further insight into the behavior for different TDOAs, we evaluate the performance for the different source positions individually in the following.
Fig.~\ref{fig:ASR-angle} shows the recognition rate for signals processed with the proposed unbiased estimators 2, 3 and 4 for the different source positions in rooms A, B and C. While dereverberation using the heuristic DOA-dependent estimator $\CDRproptwo$ yields the highest recognition rate in almost all cases, the DOA-independent estimator $\CDRpropthree$ also achieves a significant improvement over all angles. The estimator $\CDRpropfour$, while not usable for DOA $\theta=0$ due to the disappearing imaginary part of the coherence, remarkably already achieves a significantly increased recognition rate for DOAs as small as $10^\circ$, and similar recognition rates as the DOA-independent estimator for higher DOAs. In Room A, where the mismatch between the diffuse assumption and the actual reverberation coherence is significant, the estimator slightly exceeds the performance of the (on average best) estimator $\CDRproptwo$ for some positions, indicating that in some scenarios it may be of advantage to use an estimator which does not assume an isotropic noise field.

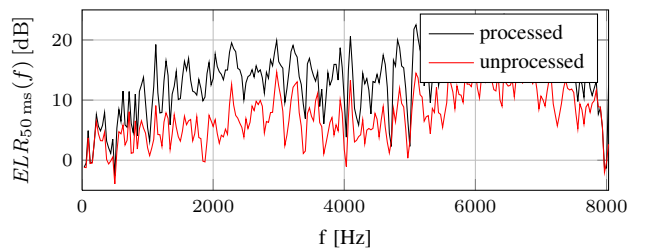
\begin{figure}[tb]
    \centering
        \setlength\figureheight{2.4cm}
	\setlength\figurewidth{7cm}
\pgfplotsset{ 
tick label style={font=\scriptsize},
label style={font=\footnotesize},
legend style={font=\footnotesize},
}
\begin{tikzpicture}

\begin{axis}[%
width=\figurewidth,
height=\figureheight,
scale only axis,
xmin=0,
xmax=8031.25,
xlabel={f [Hz]},
xmajorgrids,
ymin=-5,
ymax=25,
ylabel={$\ELR_{50\,\mathrm{ms}}(f)\,[\mathrm{dB}]$},
ymajorgrids,
legend style={draw=black,fill=white,legend cell align=left}
]
\addplot [color=black,solid]
  table[row sep=crcr]{
31.25	-1.18869912624359	\\
62.5	-0.471017688512802	\\
93.75	4.91889810562134	\\
125	-0.514225006103516	\\
156.25	-0.306524574756622	\\
187.5	3.01997995376587	\\
218.75	7.60158061981201	\\
250	5.65716695785522	\\
281.25	4.88122367858887	\\
312.5	5.80366659164429	\\
343.75	7.93434190750122	\\
375	3.18030405044556	\\
406.25	2.03581619262695	\\
437.5	1.14865565299988	\\
468.75	3.50838899612427	\\
500	-2.59677410125732	\\
531.25	5.32456111907959	\\
562.5	8.44012928009033	\\
593.75	6.46750497817993	\\
625	11.4955501556396	\\
656.25	5.8632173538208	\\
687.5	5.48483753204346	\\
718.75	11.4274997711182	\\
750	6.12825584411621	\\
781.25	4.6073694229126	\\
812.5	11.006046295166	\\
843.75	4.05444383621216	\\
875	11.4418792724609	\\
906.25	12.5346364974976	\\
937.5	13.792820930481	\\
968.75	9.50118160247803	\\
1000	5.4139232635498	\\
1031.25	3.16503882408142	\\
1062.5	8.202073097229	\\
1093.75	12.2560729980469	\\
1125	19.2471485137939	\\
1156.25	11.2362470626831	\\
1187.5	7.86199283599854	\\
1218.75	7.93567848205566	\\
1250	13.1958122253418	\\
1281.25	16.4938869476318	\\
1312.5	14.0832471847534	\\
1343.75	5.90104866027832	\\
1375	10.9953575134277	\\
1406.25	12.3086853027344	\\
1437.5	14.7567319869995	\\
1468.75	10.0922107696533	\\
1500	10.2557468414307	\\
1531.25	13.7373819351196	\\
1562.5	16.4539451599121	\\
1593.75	14.9828720092773	\\
1625	16.9334201812744	\\
1656.25	13.32337474823	\\
1687.5	11.4871587753296	\\
1718.75	11.3095436096191	\\
1750	10.829963684082	\\
1781.25	13.4210472106934	\\
1812.5	11.9885358810425	\\
1843.75	9.88722515106201	\\
1875	10.5066623687744	\\
1906.25	14.4127836227417	\\
1937.5	16.2615184783936	\\
1968.75	12.5347509384155	\\
2000	13.5853624343872	\\
2031.25	17.6568298339844	\\
2062.5	15.5824890136719	\\
2093.75	14.7242221832275	\\
2125	14.2778511047363	\\
2156.25	13.3811616897583	\\
2187.5	15.008505821228	\\
2218.75	14.9960918426514	\\
2250	17.9836311340332	\\
2281.25	19.5213527679443	\\
2312.5	18.5642642974854	\\
2343.75	18.0531196594238	\\
2375	15.2858638763428	\\
2406.25	15.042724609375	\\
2437.5	13.9097509384155	\\
2468.75	10.0961723327637	\\
2500	12.8655481338501	\\
2531.25	13.9032831192017	\\
2562.5	14.1538238525391	\\
2593.75	15.0525741577148	\\
2625	13.8406391143799	\\
2656.25	15.6953105926514	\\
2687.5	14.640323638916	\\
2718.75	15.0298519134521	\\
2750	13.7167549133301	\\
2781.25	11.1548128128052	\\
2812.5	14.631157875061	\\
2843.75	13.6257638931274	\\
2875	16.6976528167725	\\
2906.25	15.7080574035645	\\
2937.5	17.9595546722412	\\
2968.75	19.8839778900146	\\
3000	16.3982887268066	\\
3031.25	15.7799634933472	\\
3062.5	15.3115167617798	\\
3093.75	9.62264251708984	\\
3125	11.870774269104	\\
3156.25	14.7401828765869	\\
3187.5	18.0652599334717	\\
3218.75	19.0889129638672	\\
3250	17.7881469726562	\\
3281.25	17.1500606536865	\\
3312.5	13.8766107559204	\\
3343.75	12.6123914718628	\\
3375	9.32392501831055	\\
3406.25	6.92221832275391	\\
3437.5	16.0305519104004	\\
3468.75	16.9459800720215	\\
3500	15.2052927017212	\\
3531.25	14.032735824585	\\
3562.5	13.1529769897461	\\
3593.75	14.6627883911133	\\
3625	8.94545745849609	\\
3656.25	11.8295373916626	\\
3687.5	6.05518531799316	\\
3718.75	10.0760841369629	\\
3750	14.5478096008301	\\
3781.25	12.5021724700928	\\
3812.5	13.393949508667	\\
3843.75	15.9332275390625	\\
3875	15.9444961547852	\\
3906.25	13.3296003341675	\\
3937.5	18.7084121704102	\\
3968.75	15.4108390808105	\\
4000	10.035304069519	\\
4031.25	3.84178018569946	\\
4062.5	14.9692525863647	\\
4093.75	20.5799961090088	\\
4125	15.1075115203857	\\
4156.25	8.61332702636719	\\
4187.5	7.08142328262329	\\
4218.75	5.85225963592529	\\
4250	8.72506237030029	\\
4281.25	14.4604797363281	\\
4312.5	12.7147884368896	\\
4343.75	13.5702686309814	\\
4375	11.6558847427368	\\
4406.25	7.70379829406738	\\
4437.5	13.0898094177246	\\
4468.75	12.9222316741943	\\
4500	14.2681264877319	\\
4531.25	12.8880681991577	\\
4562.5	15.0000305175781	\\
4593.75	17.4723587036133	\\
4625	15.9357213973999	\\
4656.25	12.1410102844238	\\
4687.5	4.49285316467285	\\
4718.75	2.30808615684509	\\
4750	16.6406860351562	\\
4781.25	15.6659908294678	\\
4812.5	19.9940090179443	\\
4843.75	18.7357635498047	\\
4875	15.6679105758667	\\
4906.25	15.3614797592163	\\
4937.5	11.9130592346191	\\
4968.75	6.70364809036255	\\
5000	2.29855513572693	\\
5031.25	13.5894603729248	\\
5062.5	21.7851600646973	\\
5093.75	22.5121097564697	\\
5125	20.2830429077148	\\
5156.25	18.417013168335	\\
5187.5	19.340202331543	\\
5218.75	12.3645114898682	\\
5250	14.6784515380859	\\
5281.25	14.4745254516602	\\
5312.5	9.5591983795166	\\
5343.75	13.594536781311	\\
5375	18.0804386138916	\\
5406.25	19.3466606140137	\\
5437.5	12.3955821990967	\\
5468.75	14.2566404342651	\\
5500	17.782735824585	\\
5531.25	15.9534597396851	\\
5562.5	14.9987144470215	\\
5593.75	21.2766876220703	\\
5625	18.8759269714355	\\
5656.25	19.0281066894531	\\
5687.5	17.0071220397949	\\
5718.75	21.5675010681152	\\
5750	19.7130432128906	\\
5781.25	17.9639377593994	\\
5812.5	15.2088108062744	\\
5843.75	21.3354015350342	\\
5875	23.5781173706055	\\
5906.25	21.741569519043	\\
5937.5	21.3226623535156	\\
5968.75	23.0916557312012	\\
6000	22.8435859680176	\\
6031.25	20.5863571166992	\\
6062.5	17.7054996490479	\\
6093.75	20.9161491394043	\\
6125	20.6624031066895	\\
6156.25	19.6247634887695	\\
6187.5	23.415843963623	\\
6218.75	22.7255535125732	\\
6250	19.6114978790283	\\
6281.25	17.6205520629883	\\
6312.5	19.11155128479	\\
6343.75	19.5766849517822	\\
6375	18.5162391662598	\\
6406.25	17.9611873626709	\\
6437.5	24.3598899841309	\\
6468.75	18.741174697876	\\
6500	16.3954315185547	\\
6531.25	18.2214336395264	\\
6562.5	20.9412460327148	\\
6593.75	18.3490524291992	\\
6625	21.8650035858154	\\
6656.25	22.7859783172607	\\
6687.5	21.6690406799316	\\
6718.75	20.1157722473145	\\
6750	20.7958602905273	\\
6781.25	18.1541233062744	\\
6812.5	17.6057014465332	\\
6843.75	19.6729011535645	\\
6875	16.9517974853516	\\
6906.25	19.936408996582	\\
6937.5	21.5175743103027	\\
6968.75	13.7760410308838	\\
7000	16.4476356506348	\\
7031.25	20.074556350708	\\
7062.5	18.4310913085938	\\
7093.75	16.5285148620605	\\
7125	17.2011833190918	\\
7156.25	20.3473873138428	\\
7187.5	14.7388153076172	\\
7218.75	16.7198638916016	\\
7250	18.3834285736084	\\
7281.25	18.6263542175293	\\
7312.5	18.1913642883301	\\
7343.75	20.0789222717285	\\
7375	20.0466251373291	\\
7406.25	21.7431621551514	\\
7437.5	20.1519622802734	\\
7468.75	19.0048770904541	\\
7500	16.5050735473633	\\
7531.25	12.9551486968994	\\
7562.5	9.77717971801758	\\
7593.75	13.5094156265259	\\
7625	12.6708326339722	\\
7656.25	10.4587135314941	\\
7687.5	16.2992420196533	\\
7718.75	14.6092567443848	\\
7750	12.0089817047119	\\
7781.25	10.764479637146	\\
7812.5	11.9763259887695	\\
7843.75	13.1287937164307	\\
7875	7.65925407409668	\\
7906.25	8.67612648010254	\\
7937.5	5.42399120330811	\\
7968.75	-1.26603388786316	\\
8000	-1.30016779899597	\\
8031.25	12.461950302124	\\
};
\addlegendentry{processed};
\addplot [color=red]
  table[row sep=crcr]{
31.25	-1.18867528438568	\\
62.5	-1.22578155994415	\\
93.75	3.71975541114807	\\
125	-0.531992673873901	\\
156.25	-0.46667543053627	\\
187.5	3.26189875602722	\\
218.75	6.45283031463623	\\
250	4.40434837341309	\\
281.25	3.2587685585022	\\
312.5	3.23138570785522	\\
343.75	4.67121410369873	\\
375	0.111905112862587	\\
406.25	-0.621898710727692	\\
437.5	-0.131624490022659	\\
468.75	0.539302349090576	\\
500	-3.94019174575806	\\
531.25	1.41670179367065	\\
562.5	4.73797798156738	\\
593.75	4.39470958709717	\\
625	5.84809732437134	\\
656.25	2.54285311698914	\\
687.5	3.31938767433167	\\
718.75	8.07189655303955	\\
750	1.15055286884308	\\
781.25	1.24854600429535	\\
812.5	6.54676008224487	\\
843.75	1.86184692382812	\\
875	5.22146797180176	\\
906.25	6.55060005187988	\\
937.5	5.34442329406738	\\
968.75	4.50212574005127	\\
1000	2.07304692268372	\\
1031.25	0.767931997776031	\\
1062.5	1.98630332946777	\\
1093.75	3.46223449707031	\\
1125	9.09138774871826	\\
1156.25	4.27362203598022	\\
1187.5	4.24845600128174	\\
1218.75	2.34005236625671	\\
1250	6.35654211044312	\\
1281.25	7.87941265106201	\\
1312.5	6.95126533508301	\\
1343.75	1.82879114151001	\\
1375	4.33725595474243	\\
1406.25	7.56004571914673	\\
1437.5	5.4819540977478	\\
1468.75	3.90058064460754	\\
1500	3.17368221282959	\\
1531.25	4.82483291625977	\\
1562.5	7.32848358154297	\\
1593.75	5.90357780456543	\\
1625	6.55437469482422	\\
1656.25	6.78037452697754	\\
1687.5	3.22530388832092	\\
1718.75	4.37957715988159	\\
1750	7.23002529144287	\\
1781.25	3.76646280288696	\\
1812.5	3.60123896598816	\\
1843.75	-0.161215662956238	\\
1875	-0.278774678707123	\\
1906.25	3.13087916374207	\\
1937.5	7.45439004898071	\\
1968.75	6.57115840911865	\\
2000	5.62581920623779	\\
2031.25	6.74176263809204	\\
2062.5	4.44911289215088	\\
2093.75	2.82351636886597	\\
2125	4.16486692428589	\\
2156.25	3.88137221336365	\\
2187.5	5.16569423675537	\\
2218.75	4.30222082138062	\\
2250	9.33427238464355	\\
2281.25	12.6268854141235	\\
2312.5	9.18183708190918	\\
2343.75	7.45946884155273	\\
2375	6.99957942962646	\\
2406.25	6.17852926254272	\\
2437.5	5.47336101531982	\\
2468.75	3.84249091148376	\\
2500	6.89528560638428	\\
2531.25	5.2725625038147	\\
2562.5	6.29590225219727	\\
2593.75	9.01316452026367	\\
2625	8.26812553405762	\\
2656.25	9.80545711517334	\\
2687.5	9.26198101043701	\\
2718.75	9.93431282043457	\\
2750	8.99213600158691	\\
2781.25	4.51700592041016	\\
2812.5	6.43458080291748	\\
2843.75	6.29460287094116	\\
2875	7.92615032196045	\\
2906.25	7.96616744995117	\\
2937.5	11.433988571167	\\
2968.75	14.6660346984863	\\
3000	12.3443374633789	\\
3031.25	10.5608758926392	\\
3062.5	8.14232158660889	\\
3093.75	2.47009539604187	\\
3125	4.17276477813721	\\
3156.25	6.59929847717285	\\
3187.5	9.91590118408203	\\
3218.75	12.1247386932373	\\
3250	12.6163158416748	\\
3281.25	13.0381488800049	\\
3312.5	7.91401386260986	\\
3343.75	5.66555595397949	\\
3375	1.12557899951935	\\
3406.25	1.52943348884583	\\
3437.5	7.7388744354248	\\
3468.75	8.71914672851562	\\
3500	7.30892181396484	\\
3531.25	6.5368766784668	\\
3562.5	9.89354228973389	\\
3593.75	11.1136960983276	\\
3625	6.25783061981201	\\
3656.25	5.59734725952148	\\
3687.5	1.08376657962799	\\
3718.75	2.67599582672119	\\
3750	4.43936538696289	\\
3781.25	6.27241086959839	\\
3812.5	7.09427070617676	\\
3843.75	4.21004819869995	\\
3875	5.88526344299316	\\
3906.25	4.90131950378418	\\
3937.5	10.0301780700684	\\
3968.75	7.82797574996948	\\
4000	1.76938951015472	\\
4031.25	-1.05849730968475	\\
4062.5	7.04929256439209	\\
4093.75	13.2734375	\\
4125	7.31247138977051	\\
4156.25	3.03830718994141	\\
4187.5	3.30923438072205	\\
4218.75	3.92307734489441	\\
4250	4.22941017150879	\\
4281.25	6.72888040542603	\\
4312.5	2.75416612625122	\\
4343.75	5.00028562545776	\\
4375	5.28291273117065	\\
4406.25	5.05575752258301	\\
4437.5	5.29720687866211	\\
4468.75	7.77847528457642	\\
4500	6.83265161514282	\\
4531.25	3.86900615692139	\\
4562.5	6.6120343208313	\\
4593.75	8.90936279296875	\\
4625	5.51028156280518	\\
4656.25	2.99606275558472	\\
4687.5	1.20828545093536	\\
4718.75	3.25352764129639	\\
4750	8.58319664001465	\\
4781.25	6.38497114181519	\\
4812.5	9.17351150512695	\\
4843.75	11.9789199829102	\\
4875	9.44752025604248	\\
4906.25	8.85435485839844	\\
4937.5	5.1515965461731	\\
4968.75	0.340834587812424	\\
5000	3.09279799461365	\\
5031.25	8.34192085266113	\\
5062.5	11.8453893661499	\\
5093.75	14.4546699523926	\\
5125	13.7027559280396	\\
5156.25	10.9043941497803	\\
5187.5	10.5041103363037	\\
5218.75	6.49141550064087	\\
5250	7.22316646575928	\\
5281.25	7.41619968414307	\\
5312.5	4.39777421951294	\\
5343.75	8.61141967773438	\\
5375	7.95037317276001	\\
5406.25	12.1320924758911	\\
5437.5	7.90230083465576	\\
5468.75	7.63631582260132	\\
5500	5.47350454330444	\\
5531.25	7.6210823059082	\\
5562.5	9.55117607116699	\\
5593.75	15.1047763824463	\\
5625	12.7488956451416	\\
5656.25	11.355375289917	\\
5687.5	8.59557628631592	\\
5718.75	12.3542556762695	\\
5750	10.9277076721191	\\
5781.25	10.8258361816406	\\
5812.5	9.63526630401611	\\
5843.75	13.9982662200928	\\
5875	12.6446104049683	\\
5906.25	12.6633920669556	\\
5937.5	12.2466650009155	\\
5968.75	14.0659952163696	\\
6000	14.67551612854	\\
6031.25	11.9698572158813	\\
6062.5	10.3348274230957	\\
6093.75	13.9185733795166	\\
6125	15.5249671936035	\\
6156.25	13.1297779083252	\\
6187.5	14.5024213790894	\\
6218.75	15.1390914916992	\\
6250	13.2783422470093	\\
6281.25	11.7422876358032	\\
6312.5	11.5226316452026	\\
6343.75	14.5286550521851	\\
6375	10.0275049209595	\\
6406.25	8.99474811553955	\\
6437.5	15.2003307342529	\\
6468.75	12.8020286560059	\\
6500	12.329568862915	\\
6531.25	12.4556045532227	\\
6562.5	13.2704849243164	\\
6593.75	10.5311260223389	\\
6625	14.0864715576172	\\
6656.25	14.1369075775146	\\
6687.5	14.1632566452026	\\
6718.75	12.8362321853638	\\
6750	13.2008562088013	\\
6781.25	14.7381992340088	\\
6812.5	13.6825866699219	\\
6843.75	13.8361396789551	\\
6875	12.7739524841309	\\
6906.25	14.1991424560547	\\
6937.5	11.8113946914673	\\
6968.75	8.51004028320312	\\
7000	10.6535482406616	\\
7031.25	10.0328941345215	\\
7062.5	9.00482082366943	\\
7093.75	11.4284400939941	\\
7125	13.479513168335	\\
7156.25	13.66943359375	\\
7187.5	9.36727619171143	\\
7218.75	7.53026533126831	\\
7250	5.71929502487183	\\
7281.25	7.51056909561157	\\
7312.5	9.41298389434814	\\
7343.75	11.9062633514404	\\
7375	13.3573894500732	\\
7406.25	11.9610071182251	\\
7437.5	11.6866149902344	\\
7468.75	11.8375568389893	\\
7500	11.2649526596069	\\
7531.25	7.93063735961914	\\
7562.5	5.21613550186157	\\
7593.75	9.36611938476562	\\
7625	8.38724517822266	\\
7656.25	6.03003072738647	\\
7687.5	8.85293769836426	\\
7718.75	9.385422706604	\\
7750	7.97041368484497	\\
7781.25	9.467209815979	\\
7812.5	9.90658855438232	\\
7843.75	11.5834560394287	\\
7875	8.88871097564697	\\
7906.25	8.82491016387939	\\
7937.5	5.52140235900879	\\
7968.75	-2.02037143707275	\\
8000	-0.321621716022491	\\
8031.25	2.66244530677795	\\
};
\addlegendentry{unprocessed};
\end{axis}
\end{tikzpicture}%
         \vspace{-1.5mm}
    \caption{Time-averaged $\ELR_{50\,\mathrm{ms}}$ as function of frequency (room B, $l=2\,\mathrm{m}$, $d=8\,\mathrm{cm}$), for unprocessed reverberant signal, and signal dereverberated using the proposed unbiased estimator 2.}
    \label{fig:ELR_f}
\end{figure}

Fig.~\ref{fig:ELR_f} shows the time-averaged $\ELR_{50\,\mathrm{ms}}$ for different frequencies before and after processing for an exemplary scenario (room B, $l=2\,\mathrm{m}$, $d=8\,\mathrm{cm}$), where $\widehat{\CDR}_\text{prop2}$ was used for dereverberation. It can be seen that the dereverberation is most effective at frequencies above 1000\,Hz, but is already significant at frequencies as low as 300\,Hz.

\section{Conclusion}
\label{sec:conclusion}
Several well-known and some novel CDR estimation methods and their application to dereverberation have been investigated. Using simulated and measured RIRs for different environments, it has been confirmed that the commonly used model of a reverberant speech signal as a plane wave in diffuse noise is sufficiently accurate to justify the application of CDR-based signal enhancement to dereverberation. However, the known CDR estimators were found to be either biased or not robust enough for practical application to signal enhancement. It has been shown that several variants of unbiased estimators can be derived which improve robustness towards model errors, and that knowledge of either the signal DOA or the noise coherence is sufficient for estimation of the CDR. Employing the improved estimators for dereverberation has been shown to lead to improved dereverberation performance. Using the DOA-independent estimator, the proposed signal enhancement scheme constitutes a completely blind dereverberation system which requires no knowledge or estimation of the signal DOA.

\section*{Appendix: Definition of the ELR}
Reverberant microphone signals $x_i(t)$ can be written as a convolution of RIRs $h_i(t)$ with a clean signal $d(t)$, i.e., $x_i(t)=h_i(t)*d(t)$. The RIRs can be split at $t=T_\mathrm{e}$ into an early part containing direct path and early reflections, and a late part containing reverberation.
To quantify the amount of reverberation in a signal, the early-to-late power ratio $\ELR_{T_\mathrm{e}}$ can then be defined as the power ratio between the components created by convolution with the early RIR, and the reverberation components created by convolution with the late RIR, where $T_\mathrm{e}$ is set to an appropriate threshold, e.g., $T_\mathrm{e}=50\,\mathrm{ms}$ \cite{kuttruff_room_2000}. When $T_\mathrm{e}$ is set to include only the direct path in the early component, the ELR is equivalent to the DRR. 
For the evaluation in this paper, the $\ELR_{T_\mathrm{e}}$ is computed for the unprocessed microphone signals, and for the signals at the output of the signal enhancement system by processing the early and late signal components separately.
\bibliographystyle{IEEEtran}

\end{document}